\documentclass[fleqn,usenatbib]{aa}
\usepackage[T1]{fontenc}

\DeclareRobustCommand{\VAN}[3]{#2}
\let\VANthebibliography\thebibliography
\def\thebibliography{\DeclareRobustCommand{\VAN}[3]{##3}\VANthebibliography}

\usepackage{graphicx}	
\usepackage[export]{adjustbox}

\usepackage{amsmath}	
\usepackage{hyperref}
\usepackage{comment}

\usepackage{color, xcolor}

\usepackage{txfonts}
\begin{document}

   \title{Merger of massive galaxy cluster CL0238.3+2005 at z$\simeq$0.4: \\ just after  pericenter passage? }
    \titlerunning{CL0238}
    \authorrunning{Lyskova et al.}


   \author{N. Lyskova
          \inst{1,2,5}
          \and
    E. Churazov\inst{1,2}
      \and        
      I. Khabibullin\inst{3,1,2}
    \and 
    I.F.~Bikmaev\inst{5,6}
    \and
    R.A. Burenin\inst{2}
    \and
    W.R. Forman\inst{4}
    \and
    I.M.~Khamitov\inst{5,6}
    \and
    K. Rajpurohit\inst{4}
    \and
    R. Sunyaev\inst{1,2}
    \and
    C. Jones\inst{4}
    \and
    R. Kraft\inst{4}
    \and 
    I. Zaznobin\inst{2}
    \and
    M.A.~Gorbachev\inst{5,6}
    \and
    M.V.~Suslikov\inst{5,6}
    \and
    R.I.~Gumerov\inst{5,6}
    \and
    N.A.~Sakhibullin\inst{5,6}
          }

   \institute{Max Planck Institute for Astrophysics, Karl-Schwarzschild-Str. 1, D-85741 Garching, Germany
         \and
             Space Research Institute (IKI), Profsoyuznaya 84/32, Moscow 117997, Russia
         \and
             Universitäts-Sternwarte, Fakultät für Physik, Ludwig-Maximilians-Universität München, Scheinerstr.1, 81679 München, Germany
         \and
            Center for Astrophysics, Harvard \& Smithsonian, 60 Garden St, Cambridge, MA 02138, USA 
         \and
            Kazan Federal University,  Kremlevskaya Str. 18,  420008 Kazan, Russia
         \and
            Tatarstan Academy of Sciences, Bauman Str. 20, 420111, Kazan, Russia
             }

   \date{Received August 15, 1996; accepted August 16, 1996}

 
\abstract{Massive clusters of galaxies are very rare in the observable Universe. Even rarer are mergers of such clusters observed close to  pericenter passage. Here, we report on one such case: a massive ($\sim 10^{15}\,M_\odot$) and hot ($kT\sim 10\,{\rm keV}$)  cluster CL0238.3+2005 at $z\approx 0.42$. For this cluster, we combine  X-ray data from SRG/eROSITA and Chandra, optical images from DESI, and spectroscopy from BTA and RTT-150 telescopes. The X-ray and optical morphologies suggest an ongoing merger with the projected separation of subhalos of $\sim 200 \,{\rm kpc}$. The line-of-sight velocity of galaxies tentatively associated with the two merging halos differs by 2000-3000$\,{\rm km\,s^{-1}}$. We conclude that, most plausibly, the merger axis is neither close to the line of sight nor to the sky plane. We compare CL0238 with two well-known clusters MACS0416 and Bullet, and conclude that CL0238 corresponds to an intermediate phase between the pre-merging MACS0416 cluster and the post-merger Bullet cluster. Namely, this cluster has recently (only $\lesssim 0.1\,{\rm Gyr}$ ago) experienced an almost head-on merger. We argue that this "just after" system is a very rare case and an excellent target for lensing, Sunyaev-Zeldovich effect, and X-ray studies that can constrain properties ranging from dynamics of mergers to self-interacting dark matter, and plasma effects in intracluster medium that are associated with shock waves, e.g., electron-ion equilibration efficiency and relativistic particle acceleration.
  }


   \keywords{galaxy clusters               }

   \maketitle
%

\section{Introduction}

The merging of galaxy clusters is a fundamental process in the hierarchical formation of the cosmic web. Galaxy cluster mergers undergo several distinct phases, from the initial approach and core passage to the post-merger phase and eventual relaxation. Each phase is associated with distinct observational signatures, providing valuable insights into the dynamics of galaxy clusters, dark matter properties \citep[e.g., ][]{2004ApJ...606..819M}, and microphysics of the plasma, which constitutes the X-ray emitting intracluster medium \citep[see, e.g.,][for reviews]{2007PhR...443....1M, 2016JPlPh..82c5301Z}.  On the other hand, perturbations induced by mergers can complicate the use of galaxy cluster samples as cosmological probes. In particular, the cluster number density as a function of their mass and redshift sensitively depends on cosmological parameters  \cite[e.g.,][for a review]{2012ARA&A..50..353K}. However, during mergers, clusters are far from equilibrium, and most approaches for mass estimation give biased results, thus biasing the resulting cosmological parameters \citep[for instance, ][]{2008ApJ...680...17W, 2012A&A...538A..98A}.

In this work, we study properties of a massive ($\sim 10^{15}\,{\rm M}_\odot$) cluster SRGe~CL0238.3+2005 (hereafter, we use CL0238 as an acronym for this object) at $z\approx 0.4$ using X-ray and optical observations. At first glance, due to its high mass, high luminosity in X-rays ($L_x \simeq 6.6 \times 10^{44}$ erg s$^{-1}$ in the 0.5-2.0 keV band, as estimated in \citealt{2022AstL...48..702B}), and relatively regular morphology in the low-angular-resolution ROSAT and Planck data, CL0238 might 
appear as a relaxed cluster, perfectly suited for cosmological studies.
However, optical images show a very elongated chain of galaxies. As we show later, new Chandra X-ray observations with high angular resolution clearly reveal the perturbed state of the cluster. Moreover, CL0238 appears to be in a relatively short-lived merger phase - just after pericenter passage. In combination with its high mass and an "optimal" redshift, this makes CL0238  an interesting target for gravitational lensing studies on par with clusters from the HST Frontier Fields sample \citep{2017ApJ...837...97L} and for testing self-interacting dark matter models \citep[for example,][]{2004ApJ...606..819M, 2008ApJ...679.1173R}.

\cite{2022AstL...48..702B} identified the cluster CL0238 in 
the all-sky survey conducted with the eROSITA telescope onboard the Spectrum–Roentgen–Gamma (SRG) observatory \citep{2021A&A...656A.132S, 2021A&A...647A...1P} and conducted the first spectroscopic observations of the cluster at the 6-m Big Telescope Azimuthal (BTA) telescope in the mode of long-slit spectroscopy. \cite{2022AstL...48..702B} measured the cluster redshift  $z = 0.4205$ and estimated its mass $M_{500} \simeq 9\,\times 10^{14} M_{\odot} $ from the X-ray luminosity. Historically, the cluster CL0238 was first detected as an extended X-ray source in the ROSAT All-Sky Survey Faint Source Catalogue  \citep{2000IAUC.7432....3V}  in the 0.1-2.4 keV energy band. Later,
\cite{2012ApJS..199...34W} found a prominent concentration of galaxies in the same area in the Sloan Digital Sky Survey III and assigned it to a cluster with a photometric redshift $z_{\rm phot}\approx 0.4266$, and a mass\footnote{Throughout the paper, the cluster mass is defined as $M_{\rm \Delta}  = \Delta \rho_c  4/3 \pi R^3_{\rm \Delta}$, where $R_{\rm \Delta}$ is the radius enclosing an overdensity $\Delta$  with respect to the critical density $\rho_c$ of the Universe at a given redshift.}  $M_{500} \sim 4\,\times 10^{14} M_{\odot}$  from the total luminosity of cluster member candidates within the radius of 1 Mpc\footnote{to convert $M_{200} \simeq 6\,\times 10^{14} M_{\odot}$ from \cite{2012ApJS..199...34W} to $M_{500}$ we assume the mass-concentration relation from \cite{2019ApJ...871..168D} and the Navarro-Frenk-White cluster density profile. } . 
The cluster also enters the CompRASS catalog \citep{2019yCat..36260007T} which is an all-sky catalog of galaxy clusters obtained from the joint analysis of the Planck satellite data and the ROSAT all-sky survey. According to \cite{2019yCat..36260007T}, the CL0238 mass is estimated to be $M_{500}^{\rm XSZ}=6.5^{+1.2}_{-1.3} \,\times 10^{14} M_{\odot}$.

Here, we analyze SRG/eROSITA, Chandra, and optical observations to shed light on the properties of  CL0238.  We adopt a $\Lambda$ cold dark matter cosmology with $\Omega_{\rm M} = 0.3, \Omega_{\rm \Lambda} =0.7$ and $H_0 = 70$ km s$^{-1}$ Mpc$^{-1}$. At the cluster redshift $z = 0.4205$, 1 arcmin corresponds to 332.2 kpc.

\section{Optical observations  }
\label{sec:redshifts}
Spectroscopic observations of the brightest red sequence galaxies of the SRGe CL0238.3+2005 cluster were carried out with (1) the 6-m BTA telescope \citep{2022AstL...48..702B, 2023AstL...49..599Z} and (2) the Russian-Turkish 1.5 m telescope (RTT-150) at the T\"UB\'ITAK Observatory. Using the multimode SCORPIO and SCORPIO-2 spectrographs \citep{2005AstL...31..194A, 2011BaltA..20..363A}, \cite{2022AstL...48..702B} obtained spectra for two bright galaxies in SRGe CL0238.3+2005 (marked with squares in Fig.~\ref{fig:z}) and measured their spectroscopic redshifts (see Table~\ref{tab:redshifts}). 

The RTT-150 observations were carried out in March 07-11, 2024, with the TFOSC instrument and the Andor iKon-L 936 BEX2-DD-9ZQ CCD with a size of $2048\times 2048$ pixels,  thermoelectrically cooled to - 80$^{\circ}$C. The field of view, in direct image mode, is 11$\times$11\ arcmin$^2$ with the scale of 0.33 arcsec/pixel at binning 1$\times$1.
For each observed galaxy, 
one spectrum with an exposure time of 3600 sec was obtained by using grizm 15 and a 134$\mu$ entrance slit (corresponding to 2.4 arcsec at the sky). The wavelength range is 3900 - 8900 \r{A}, and the spectral resolution is 15 \r{A} (binning 2$\times$2 was used in spectral observations).

 The redshifts of observed galaxies were determined by cross-correlation with an elliptical galaxy template spectrum\footnote{Processing of RTT-150 spectra was performed using the software developed by Irek Khamitov and Rodion Burenin and the software package \text{IRAF} (\url{https://iraf-community.github.io})}.
A detailed description can be found in  \citep{2020AstL...46....1K}. Obtained spectroscopic measurements are presented in Table~\ref{tab:redshifts} (see also \S\ref{sec:merger} and Figure~\ref{fig:z} later in the text).

\begin{table}
\caption{Spectroscopic redshifts of individual galaxies in CL0238 (measured with BTA (marked as [1]) and RTT-150 ([2])) and their line-of-sight velocities with respect to the mean cluster redshift $\langle z\rangle = 0.4191$.
} 
\label{tab:redshifts}
\centering
\begin{tabular}{  cccc }
\hline
 (RA, DEC) & z &   $V_{los} $, km s$^{-1}$       \\ \hline
(39.5862, +20.1052)  & $0.4252 \pm 0.0004$ & 1289 $\pm$ 85 & [1] \\
 Same galaxy as above & $0.4259 \pm 0.0003$  & 1437 $\pm$ 63 &[2] \\
(39.5848, +20.1057) & $0.4228 \pm 0.0004$ &  782 $\pm$ 85 &[2]   \\
 (39.5862, +20.0940) & $0.4158 \pm 0.0006$ & -697 $\pm$ 127 &[1] \\
 (39.5852, +20.0980) & $0.4104 \pm 0.0004$ & -1838  $\pm$ 85 &[2]   \\
 (39.5852, +20.0900) & $0.4213 \pm 0.0003$ & 465 $\pm$ 63 &[2]   \\
\hline 
\end{tabular}
\end{table}

\section{X-ray observations and data reduction}

CL0238 was observed by the Chandra Advanced CCD
Imaging Spectrometer (ACIS) four times (see Table~\ref{tab:chandra_obsid} for details). Observations were processed with the standard Chandra data reduction  (CXC software v. 10.12.2; CIAO v. 4.7) and calibration software
(CalDB v. 4.10.8).  Data analysis steps are described in detail in  \cite{2009ApJ...692.1033V}  and include high background period filtering, application of the latest calibration corrections to the detected X-ray photons, and determination of the background intensity in each observation. 
For spectral analysis, we generated the spectral response files that combine the position-dependent ACIS calibration with the weights proportional to the observed brightness.
The total filtered exposure time is $\approx 40$~ks. 

\begin{table}
\centering
\renewcommand{\arraystretch}{1.1}
\caption{Details of the Chandra observations of CL0238. }
\begin{tabular}{cccc}
\hline
ObsID & Instrument & Mode & Exposure, ks        \\ \hline
27404 & ACIS-I & VFAINT & 9.95 \\
29048 & ACIS-I & VFAINT & 9.96 \\
29049 & ACIS-I & VFAINT &  9.95   \\
29050 &	ACIS-I & VFAINT &	12.22 \\
\hline 
\end{tabular}
\label{tab:chandra_obsid}
\end{table}

We use the data from the eROSITA telescope \citep{2021A&A...647A...1P} on board the  \textit{SRG} mission  \citep{2021A&A...656A.132S}, launched in 2019 which started to perform the all-sky survey mission in December 2019. We use the data accumulated over four consecutive scans, with the total effective exposure amounting to $\approx 850$ seconds per point. 
Initial reduction and processing of the data were performed using standard routines of the \texttt{eSASS} software \citep{2018SPIE10699E..5GB,2021A&A...647A...1P}, while the imaging and spectral analysis were carried out with the background models, vignetting, point spread function (PSF) and spectral response function calibrations built upon the standard ones via slight modifications motivated by results of calibration and performance verification observations \citep[e.g.][]{2021A&A...651A..41C,2023MNRAS.521.5536K}.

\section{X-ray imaging and spectral analysis}
\subsection{Global view}
\label{sec:global}

\begin{figure*}
\includegraphics[angle=0,clip=true,width=\columnwidth]{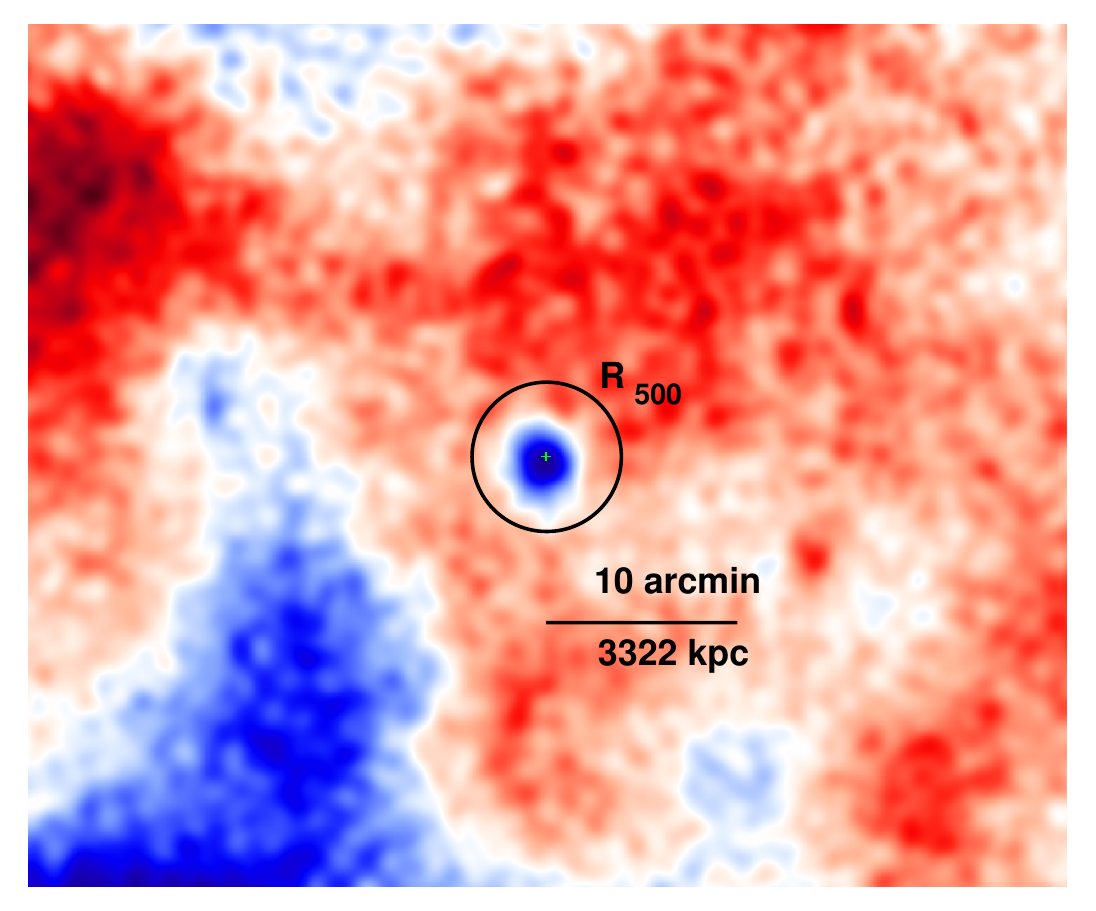}
\includegraphics[angle=0,clip=true,width=\columnwidth]{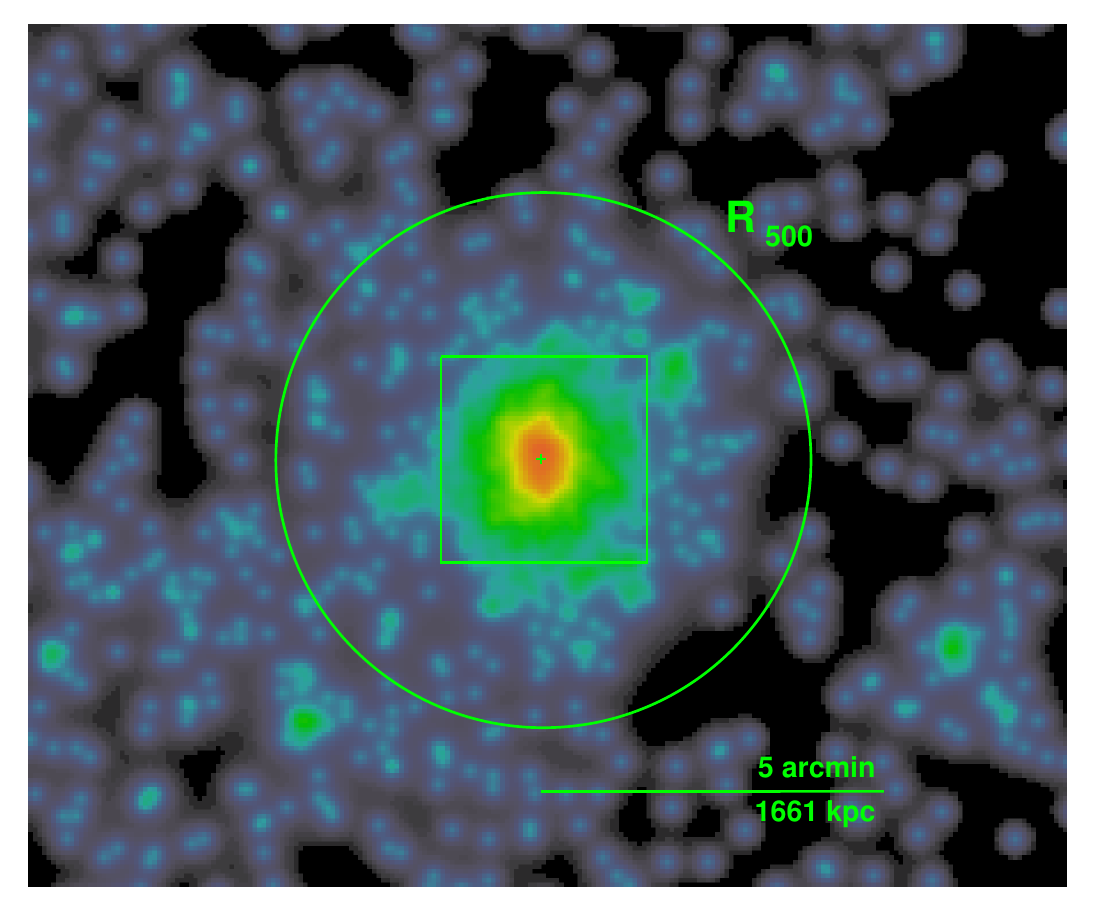} \\
\includegraphics[angle=0,width=\columnwidth]{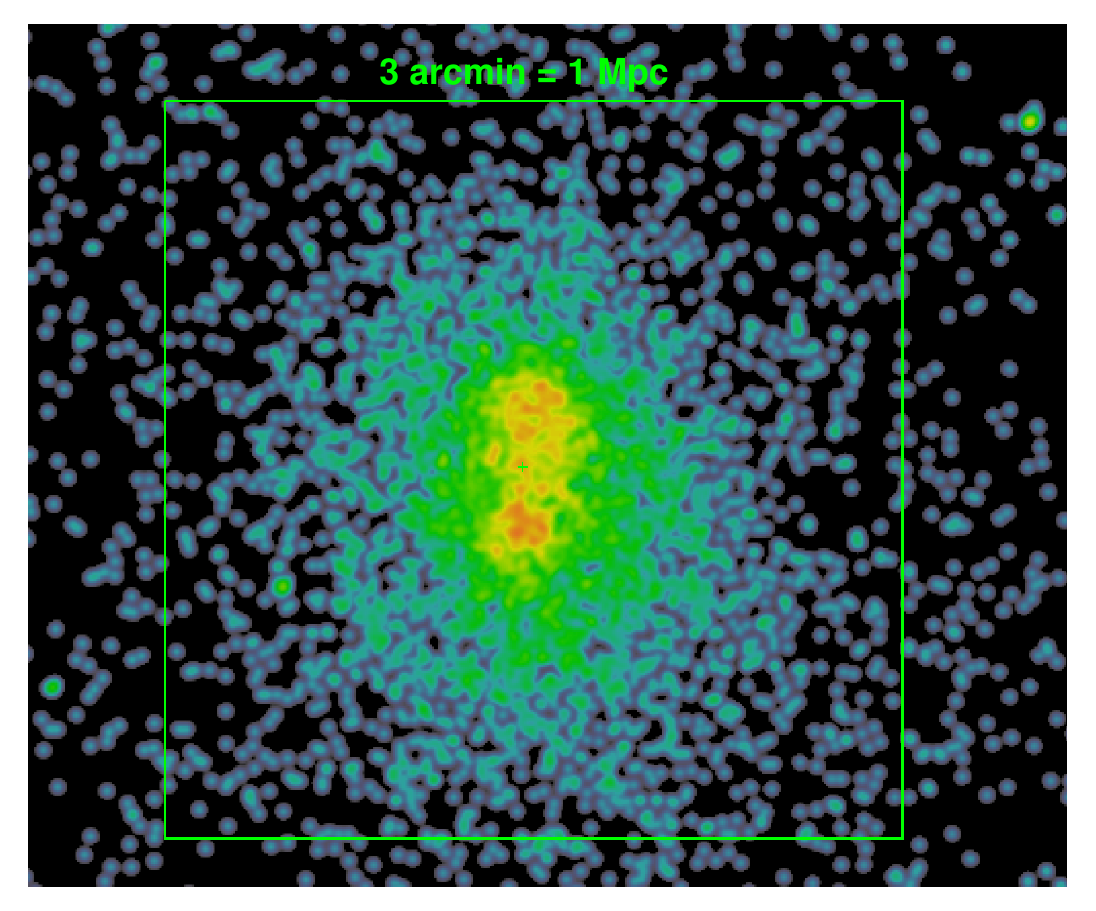}
\includegraphics[angle=0,width=\columnwidth]{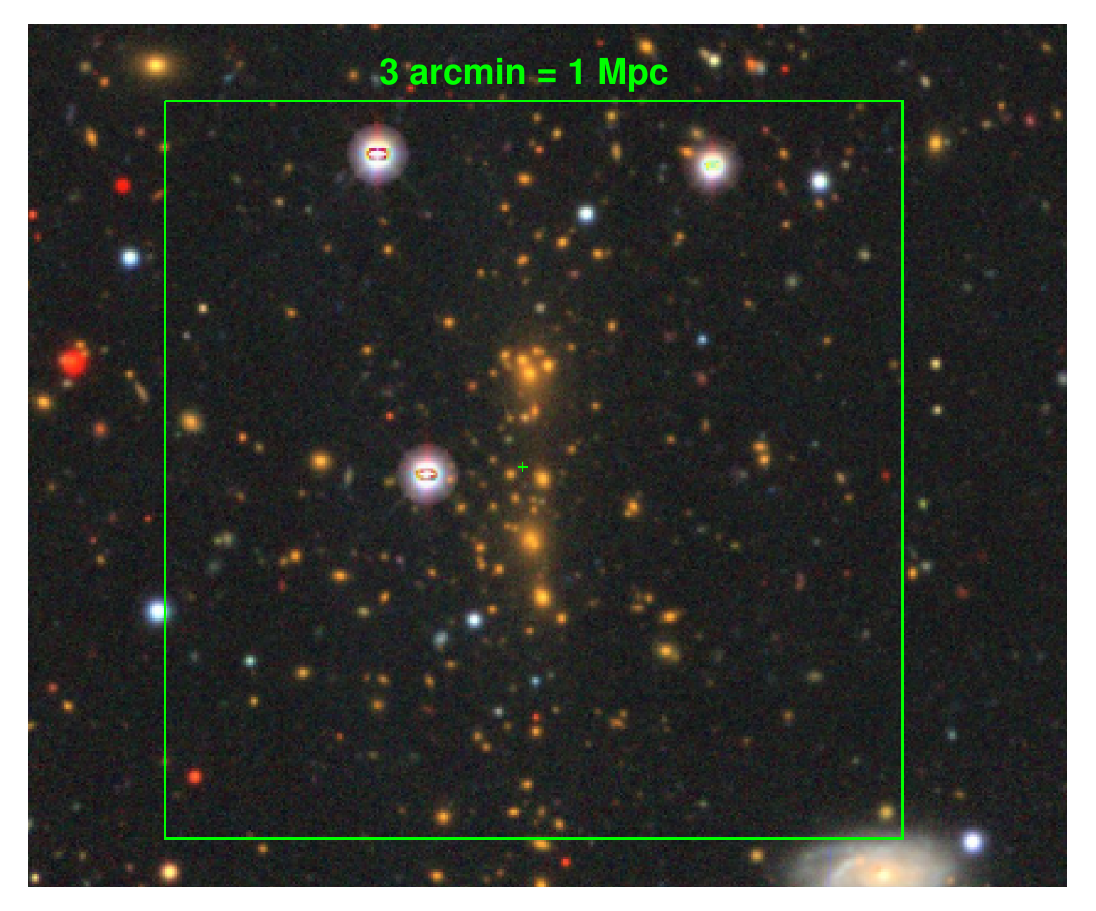}
\caption{Multiwavelength view of  CL0238. The top row shows relatively large-scale ACT (90 GHz; smoothed with 30 arcsec) and eROSITA images of the cluster in the 0.3-2.3 keV energy band. The bottom row shows the Chandra image ([0.8-4.0] keV)and the pseudo-color DESI Legacy Imaging Surveys image in the zrg (RGB) filters. Note that red galaxies at the image center are distributed along the vertical direction, and in all other images, the cluster is clearly elongated in the same direction. The observed elongation is an indication of a perturbed dynamical state of the cluster. The circle has a radius of $3.9$ arcmin  $\approx R_{500}$, the green box has a side of 3 arcmin, the central cross (RA = 02:38:20.8, DEC = +20:05:56)  marks the cluster ‘center’ defined for presentation purposes only, i.e. there is no X-ray peak at this position.}
\label{fig:review}
\end{figure*}

\begin{figure}

\includegraphics[angle=0,width=\columnwidth]{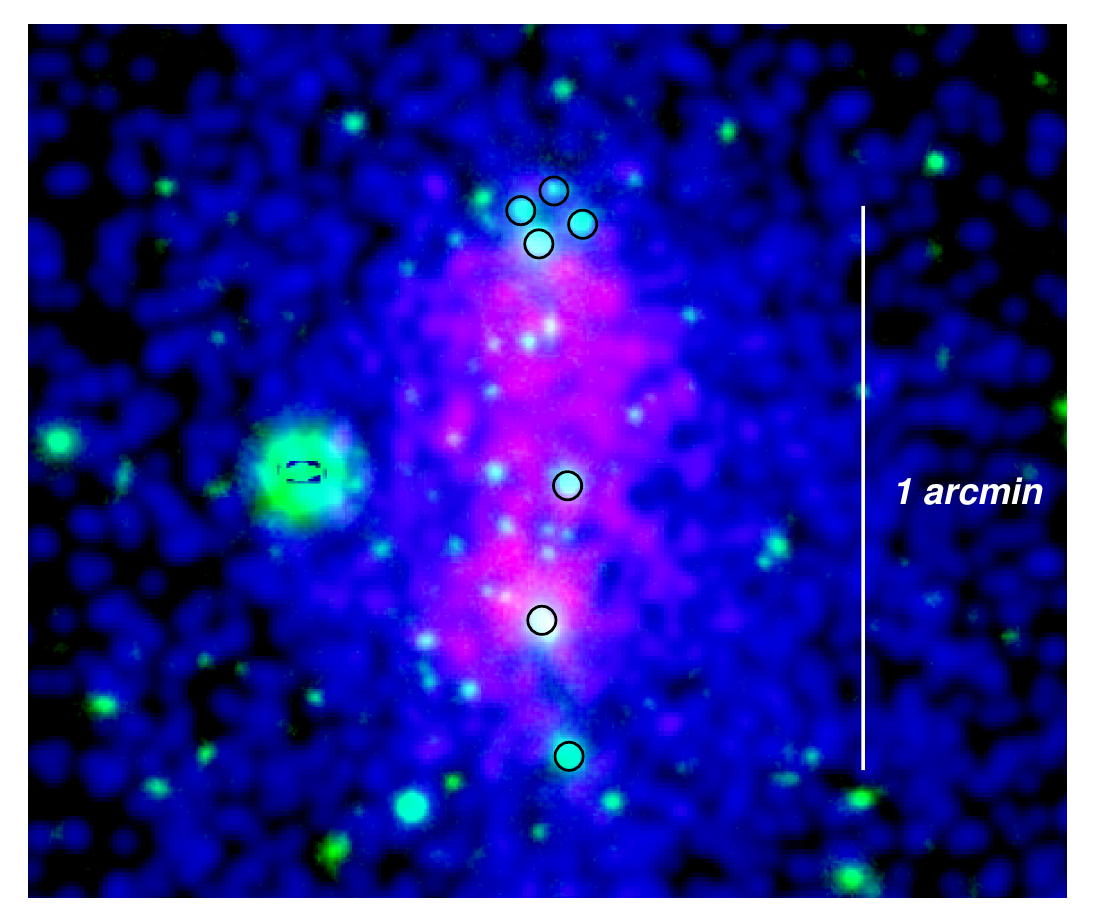}
\caption{The central part of CL0238 at optical and X-ray wavelengths. The image shows a combination of X-ray data (magenta-blue) and optical r-band DESI image  (green). A few bright galaxies that most likely belong to the cluster are highlighted with circles. Notably, a compact "group" of four visually clustered galaxies in the North lies just beyond the brightest X-ray region of the cluster.}
\label{fig:centre}
\end{figure}

Figure~\ref{fig:review} shows the cluster images at different wavelengths (see also Appendix~\ref{sec:radio} for radio data). 
In Fig.~\ref{fig:review}, the top row illustrates CL0238 at relatively large scales: 90 GHz Atacama Cosmology Telescope (ACT) image \citep[based on][]{2020JCAP...12..046N} and the eROSITA image (background subtracted, exposure corrected) in the 0.3–2.3 keV band. A small cross at the center of each image (RA = 02:38:20.8, DEC = +20:05:56)  marks the cluster `center' defined for presentation purposes from visual inspection of the eROSITA image\footnote{There is no X-ray peak at this position. Rather, the defined 'center' reflects the X-ray surface brightness distribution on scales of $1-3$ arcminutes.}. 
The cluster size, $R_{500}$\footnote{$R_{500} = \left( \frac{M_{500}}{4/3\pi \cdot 500\rho_c(z)}\right)^{1/3}$ with $M_{500}$ defined from the mass-luminosity relation in Section~\ref{sec:global.mass}}, is shown with a circle. 
The part of the cluster marked with a green box is shown in the bottom rows.
The bottom left panel shows the background subtracted, exposure, and vignetting corrected, Chandra image in the 0.8-4.0 keV energy band. The image is smoothed with the 1 arcsecond (sigma) Gaussian.  
With  Chandra's high angular resolution, the central region of CL0238 does not appear "smooth" but is resolved into clear substructures. Two bright X-ray clumps separated by  $\simeq 0.5$ arcmin $\simeq 165$ kpc (in projection)  and a `dip' ($\sim 30-40$\% dimmer relative to the clumps) between them can be seen. Such distinctive X-ray surface brightness distribution provides clear evidence that the cluster is in a perturbed dynamical state. The optical image (bottom right panel of Figure~\ref{fig:review})  exhibits a thread-like arrangement of red galaxies which is rather atypical for relaxed clusters and can be interpreted as an additional signature of the ongoing merger. Further signatures of the merger can be seen in Figure~\ref{fig:centre}, which shows the composite optical+X-ray  image of CL0238. 
The DESI r-band image is shown in green, while Chandra data are colored blue and magenta. The galaxies robustly associated with the cluster (see also redshift measurements Section~\ref{sec:redshifts}) are highlighted with black circles. A group of four galaxies to the North lies just beyond the bright X-ray clump. This picture resembles the Bullet cluster in which the hot X-ray emitting plasma lags behind the subcluster galaxies \citep{2004ApJ...606..819M}.    


\subsection{Global observables and mass estimates of CL0238}
\label{sec:global.mass}

\begin{figure}
\includegraphics[angle=0,width=\columnwidth]{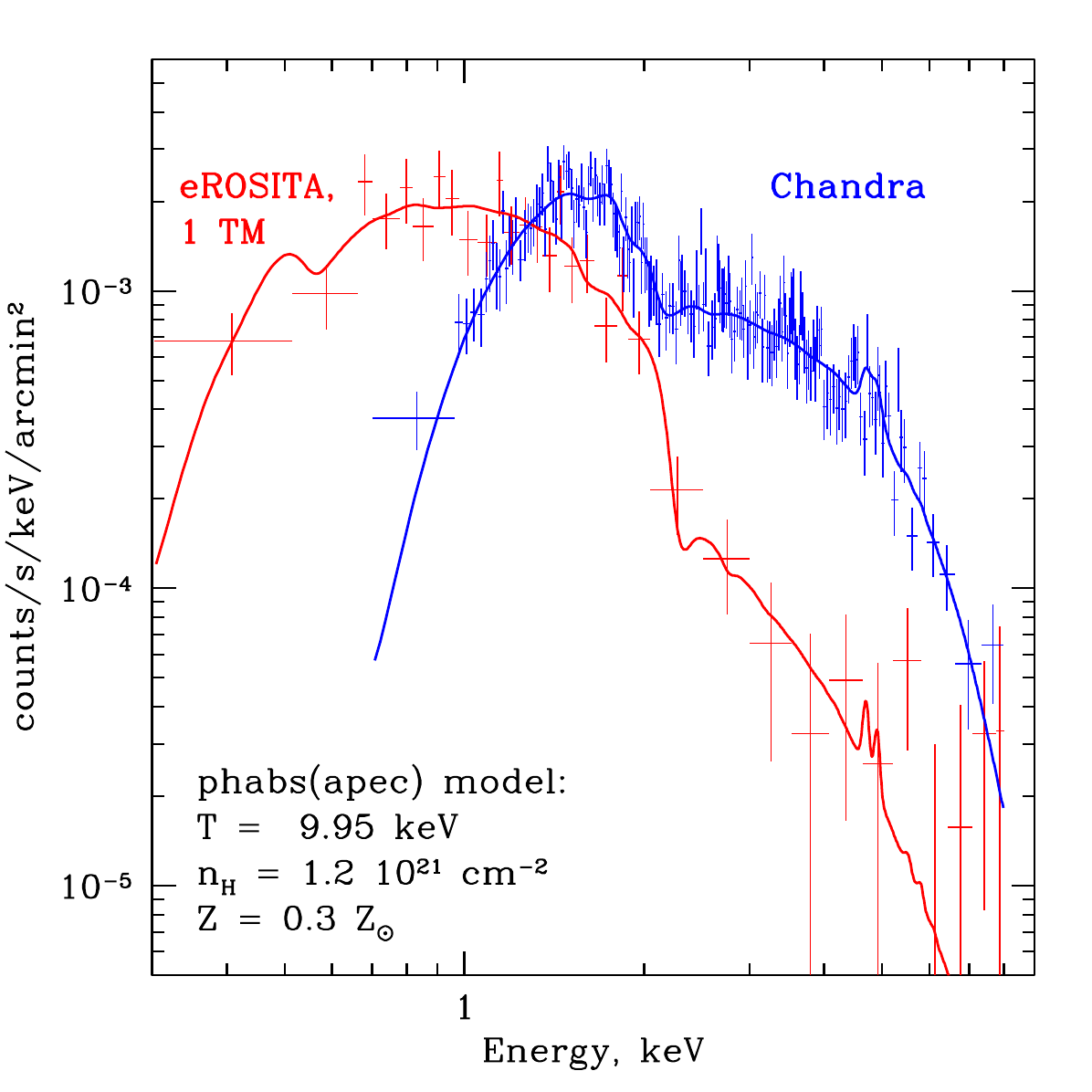}
\caption{eROSITA and Chandra spectra of CL0238.3 extracted from a circle with the  radius of $R_{500}$=$3.9$ arcmin. An annulus with $R_{500} < R <  2R_{500}$ is used as a background region. Point sources are subtracted. The eROSITA spectrum is normalized per one (out of seven) eROSITA telescope modules. The best-fit absorbed APEC model obtained from the joint analysis of eROSITA+Chandra data is shown with solid curves. 
} 
\label{fig:spectra500}
\end{figure}

To gain further insight into the cluster dynamical state, we analyze Chandra and eROSITA spectra extracted from a circular region of radius $R_{500} = 3.9$ arcmin $\simeq 1.3$ Mpc (Figure~\ref{fig:review}). For a background region, we used a ring with $R_{500} < R <  2R_{500}$. The resulting eROSITA and Chandra spectra are shown in Figure~\ref{fig:spectra500}. We fit the spectra assuming an absorbed APEC \citep{2001ApJ...556L..91S} model (\verb #phabs*apec#) with the following model parameters: the hydrogen column density set to the value of $N_H = 1.2 \cdot 10^{21}$~cm$^{-2}$ \citep[based on the approach of][]{2013MNRAS.431..394W}; the metal abundance set to $0.3$ solar; and the redshift set to $z=0.4205$. Fitting eROSITA and Chandra spectra simultaneously results in the best-fit temperature of $T_{500} = (9.95 \pm 0.79)$ keV.
The rest-frame cluster luminosity in the $[0.5,2.0]$ keV band is $L_{500} \approx 7.9 \cdot 10^{44}$ erg/s, 
and weakly sensitive to the exact value of $R_{500}$. 

Using scaling relations from \cite{2009ApJ...692.1033V} for the measured temperature and X-ray luminosity, we find 
the following estimates of the total mass: $M_{500}^{\rm L-based}\approx 9.7\times 10^{14}\,M_\odot$ and $M_{500}^{\rm T-based}\approx 9.9\times 10^{14}\,M_\odot$, respectively.
We note here, that given the ongoing merger, cluster gas is likely to be heated by a shock wave, and our X-ray-based temperature, luminosity, and mass estimate might be biased high. There are several studies where the accuracy of various ways of estimating mass was evaluated \citep[for instance, ][]{1996ApJS..104....1P, 2009ApJ...699.1004Z, 2012MNRAS.419.1766K}. A general agreement is that the accuracy of mass estimation depends on the merger geometry, its phase, and the line-of-sight angle. Therefore, the above estimate is likely subject to systematic uncertainties, which might be reduced after detailed modeling.

We conducted optical spectroscopic observations  (see Section~\ref{sec:redshifts}) of several galaxies,
which are likely to be CL0238 cluster members, 
and with these data in hand, we derive a rough estimate of the velocity dispersion of the cluster and use it as an additional mass proxy. 
Based on spectroscopic redshifts available for five galaxies in the cluster (see Table~\ref{tab:redshifts} and Fig.~\ref{fig:z}), we estimate the average redshift $\langle z \rangle = 0.4191$  and  relative line-of-sight velocities $V_{\rm los}$ of galaxies as
\begin{equation}
    \frac{V_{\rm los}}{c} = \frac{z - \langle z \rangle}{1+\langle z \rangle}.
\end{equation}
The cluster velocity dispersion can be estimated as
\begin{equation}
    \sigma = \sqrt{\frac{1}{N_{\rm gal}-1} \sum_{i=1}^{N_{\rm gal}} V_{{\rm los}, i}^2} = 1260 \,\mathrm{km s^{-1}},
\end{equation}
where $N_{\rm gal} = 5$. Taken at face value, the derived velocity dispersion in excess of $1000\,{\rm km\,s^{-1}}$ also points towards a high cluster mass \citep[e.g.][]{2008ApJ...672..122E}, although the uncertainties are large.



Even though all our estimates are subject to different biases due to the perturbed state of the cluster, they are all roughly consistent with each other and indicate a high cluster mass.  

\subsection{Signatures of a disrupted cool core?}

\begin{figure*}
\includegraphics[angle=0,width=\columnwidth]{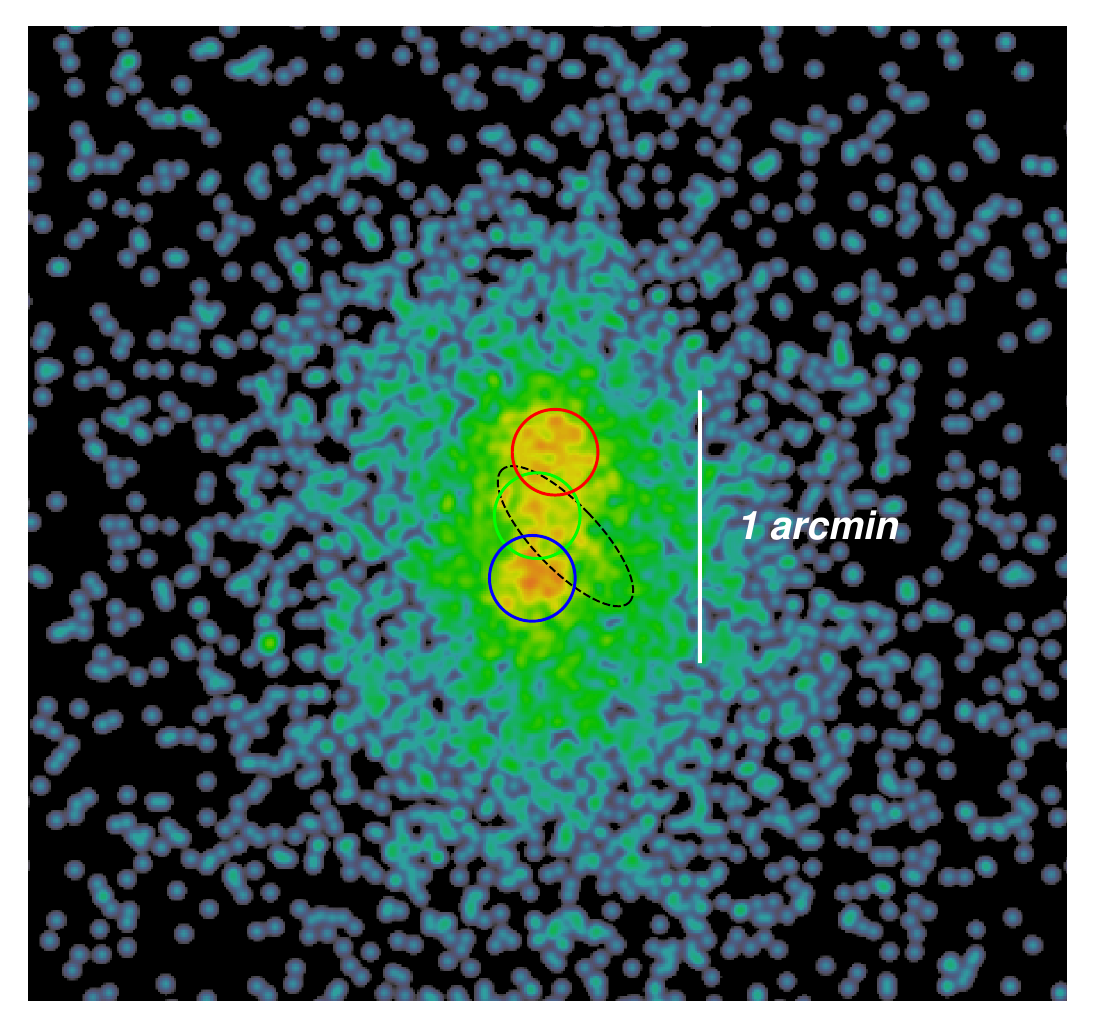}
\includegraphics[angle=0,width=\columnwidth]{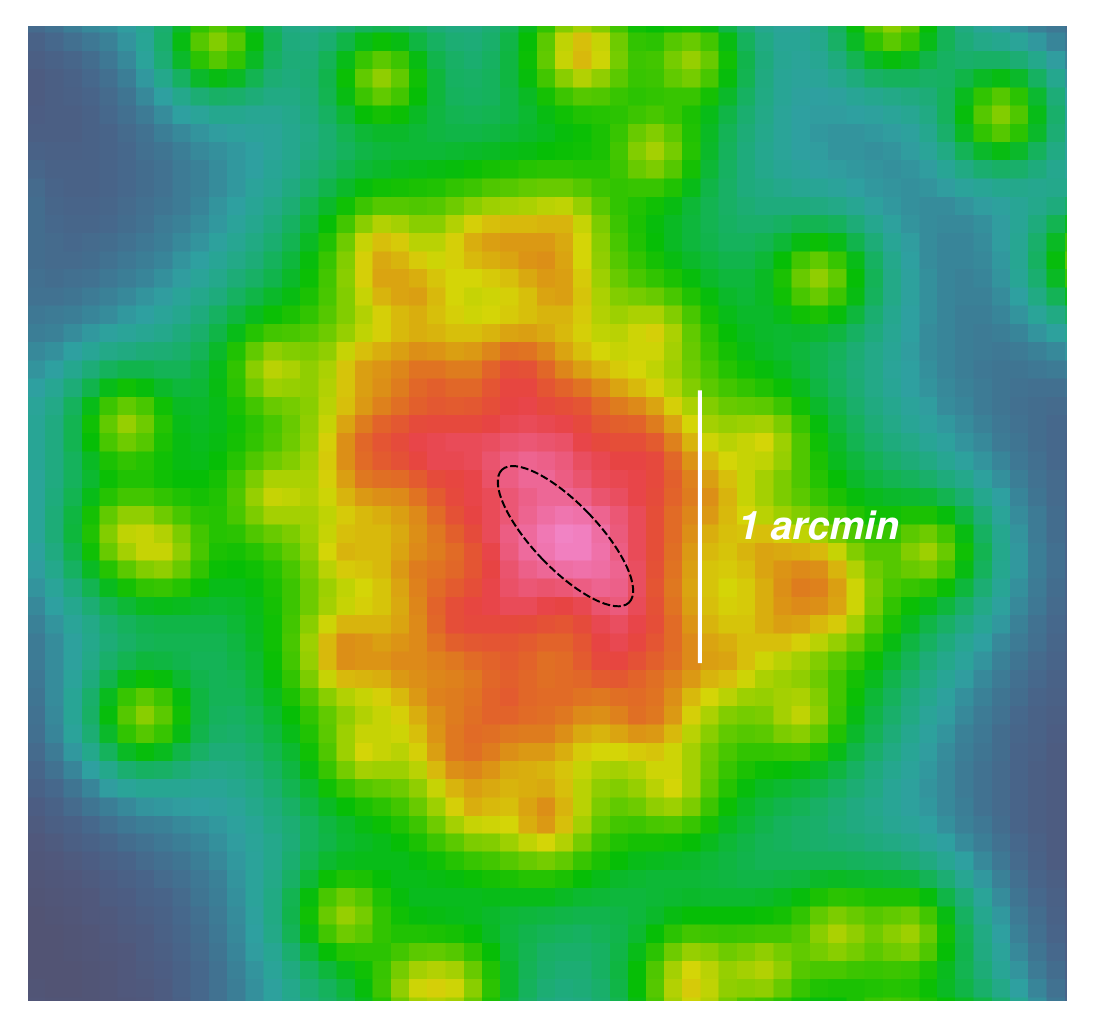}
\caption{{\bf Left:} Chandra X-ray image with three circular regions ($\simeq$ 19 arcsec in diameter) used to analyze projected spectra. The best-fitting temperatures are $10.2\pm2\,{\rm keV}$ for the bright region to the North (red circle) and $10\pm 1.9\,{\rm keV}$ for the bright region to the South (blue circle). The temperature measured for the central region (green circle) is slightly lower $7.9\pm1.4\,{\rm keV}$, although the uncertainties of all three measurements are large and the assumption that the temperatures are consistent cannot be rejected. {\bf Right:} eROSITA image in the $[0.3;0.7]$ keV energy range, where the ACIS efficiency drops significantly (see Fig. 6.8 in    
\href{https://cxc.harvard.edu/proposer/POG/html/chap6.html\#tth_sEc6.2)}{\texttt{The Chandra Proposers' Observatory Guide)}}. The region of slightly enhanced surface brightness is marked with the dashed ellipse. 
This ellipse lies just in between the North and South regions with high X-ray surface brightness in the Chandra image, corroborating the presence of colder gas in the region. We speculate that this gas might be stripped from the core of the merging/merged sub-clusters. 
}
\label{fig:clumps}
\end{figure*}

As seen from the Chandra X-ray images of CL0238, the cluster can be visually divided into two bright areas to the North and to the South, and a `dip' in between. We extract spectra from these three regions (shown as circles of different colors in the top left panel of Figure~\ref{fig:clumps}) and estimate the gas temperature. The Northern and Southern regions have a temperature of $T\simeq 10\pm 2$ keV while the region in the middle is colder, $T \simeq 8\pm 1.4$ keV. 
The uncertainties on derived temperatures are large, and, therefore, the evidence for the colder region is marginal. To make an additional test for the presence of colder gas, we use the eROSITA image in the soft energy band 0.3-0.7 keV (right panel of Figure~\ref{fig:clumps}). A tentative excess of soft photons is seen at the cluster center. The region with sightly enhanced surface brightness in the 0.3-0.7 keV band is marked with the black ellipse. This ellipse overlaps with the middle region with colder gas as measured using Chandra data. Despite these hints, the evidence for the cool gas is marginal and more data are needed to confirm it. 

Cold gas at a cluster center is often observed for 'cool core' clusters, having the peak of the X-ray surface brightness and the lowest gas temperatures centered at the brightest elliptical galaxy. This is clearly not the case for CL0238. Indeed, the brightest galaxies in the most prominent Northern group are shifted from the peak of X-ray emission and the cool gas (if any) is even farther away from these galaxies.    From a simple $\beta$-model approximation\footnote{with $\beta = 0.69$, $r_c = 0.47$ arcmin and the cluster center at RA = 02:38:20.8, Dec = +20:05:56} of the cluster radial profile and an APEC spectral model, we estimate the central electron density in the cluster core, $n_e\approx 2\times 10^{-3}\,\rm cm^{-3}$. Adopting a cooling function of \cite{1993ApJS...88..253S} with the abundance of heavy elements $1/3$ solar, we obtain the cooling time of the gas $\sim 1.6\times 10^{10}\, {\rm yr}$, i.e. almost twice longer than the Hubble time at $z=0.42$.

The derived long cooling time, combined with the lack of bright optical galaxies in this region and the absence of a prominent peak in X-rays, suggests that the softer emission is not a canonical cool core. Instead, it could be stripped gas from a cool-core which one of the merging sub-halos possessed before the merger.

\section{Merger kinematics}
\label{sec:merger}
\begin{figure*}
\includegraphics[angle=0,width=\columnwidth]{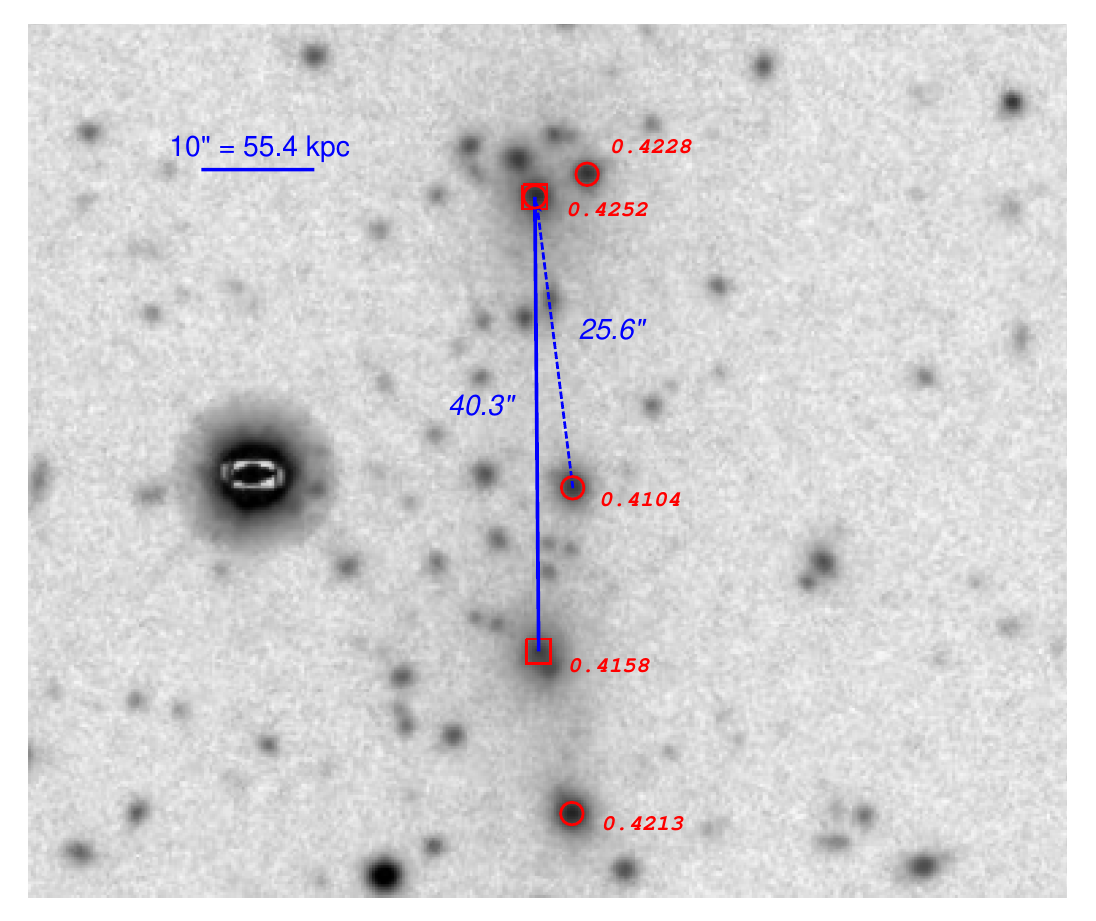}
\includegraphics[angle=0,width=\columnwidth]{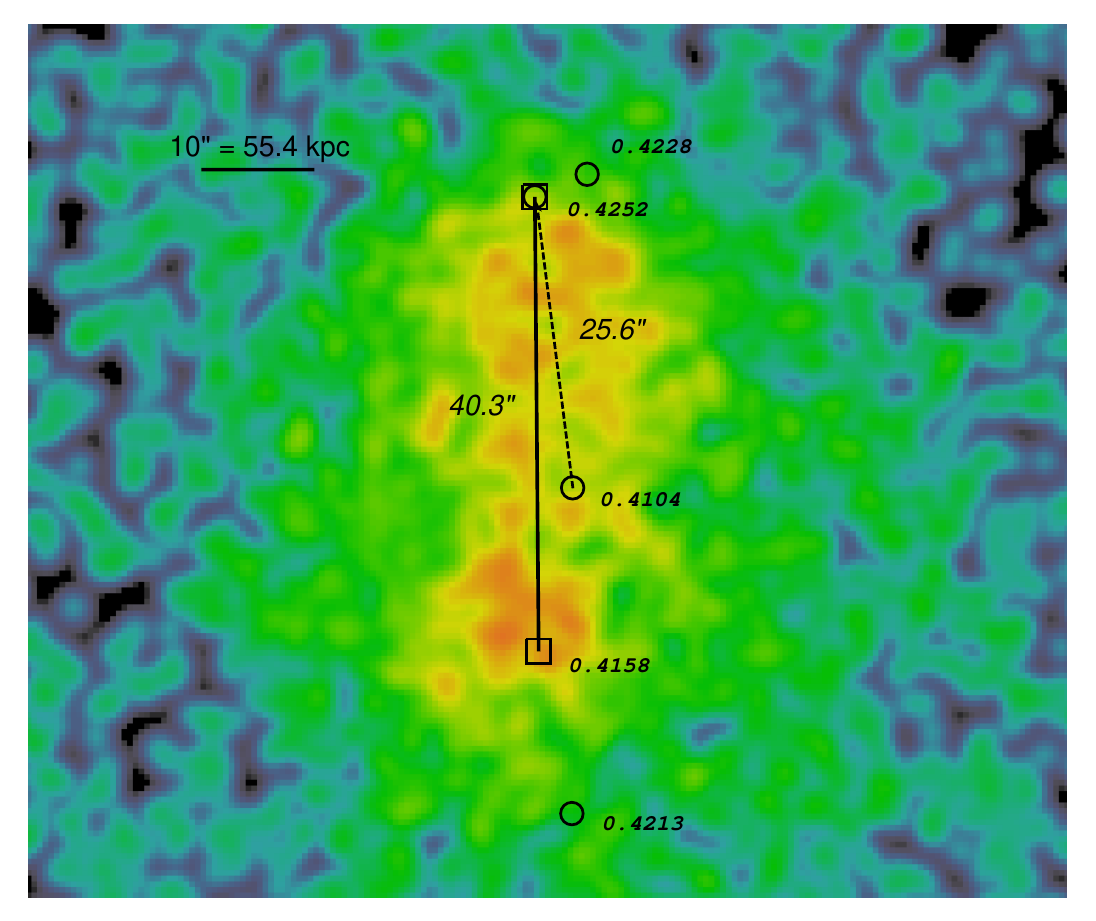}
\caption{Spectroscopic redshifts of individual galaxies in SRGe CL0238.3+2005. Galaxies observed with the 6-m BTA telescope are marked as squares, and those observed with the 1.5 Russian-Turkish telescope are marked with circles. 
Measured redshifts (see Table~\ref{tab:redshifts}) are presented next to symbols (a square or a circle). The left panel is the r-band DESI image and the right panel is the Chandra image. One subhalo is plausibly associated with four visually clustered galaxies to the North. For the second subhalo, we consider two options: a galaxy with $z=0.4158$ which lies within the southern bright X-ray region (see the right panel), and a galaxy with  $z=0.4104$ which lies between bright X-ray peaks.} 
\label{fig:z}
\end{figure*}

\subsection{Merger geometry and velocity}

\begin{figure*}
\includegraphics[angle=0,trim=1cm 5.5cm 1cm 2.5cm,clip,width=\columnwidth]{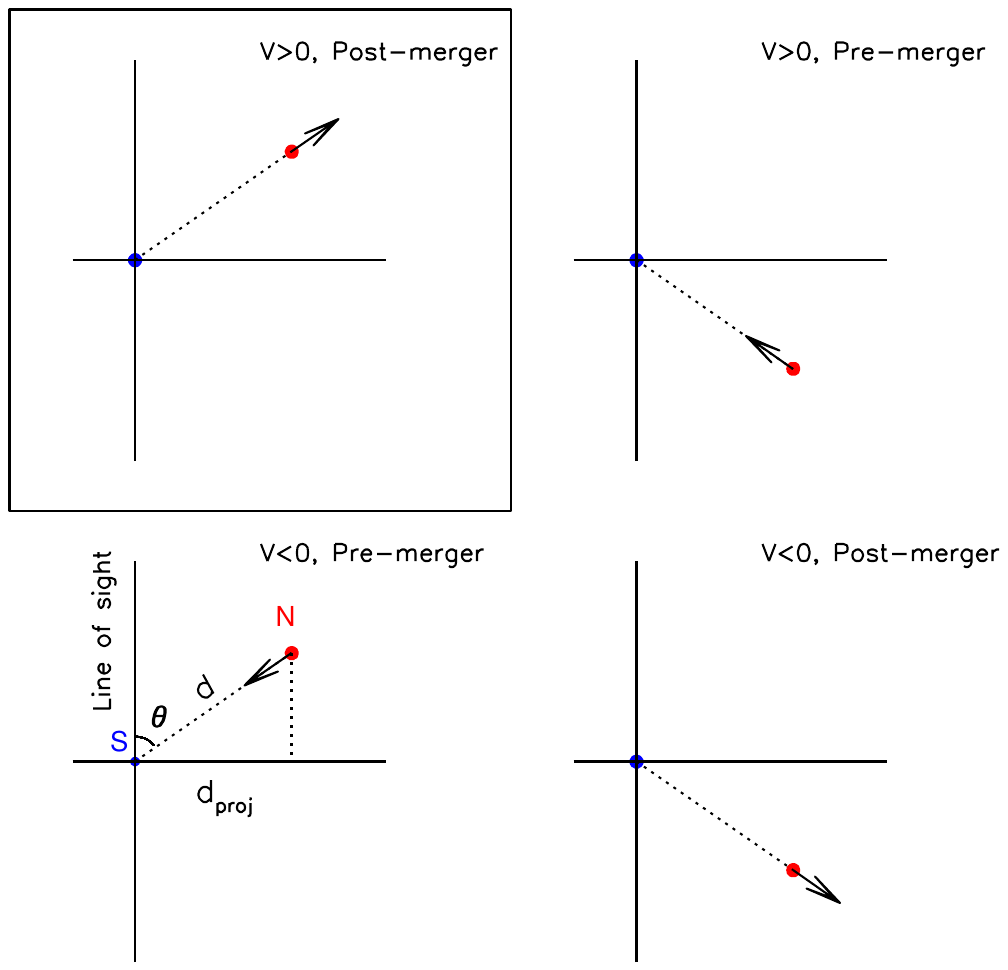}
\includegraphics[angle=0,trim=0.5cm 5.0cm 2cm 2.5cm,clip,width=\columnwidth]{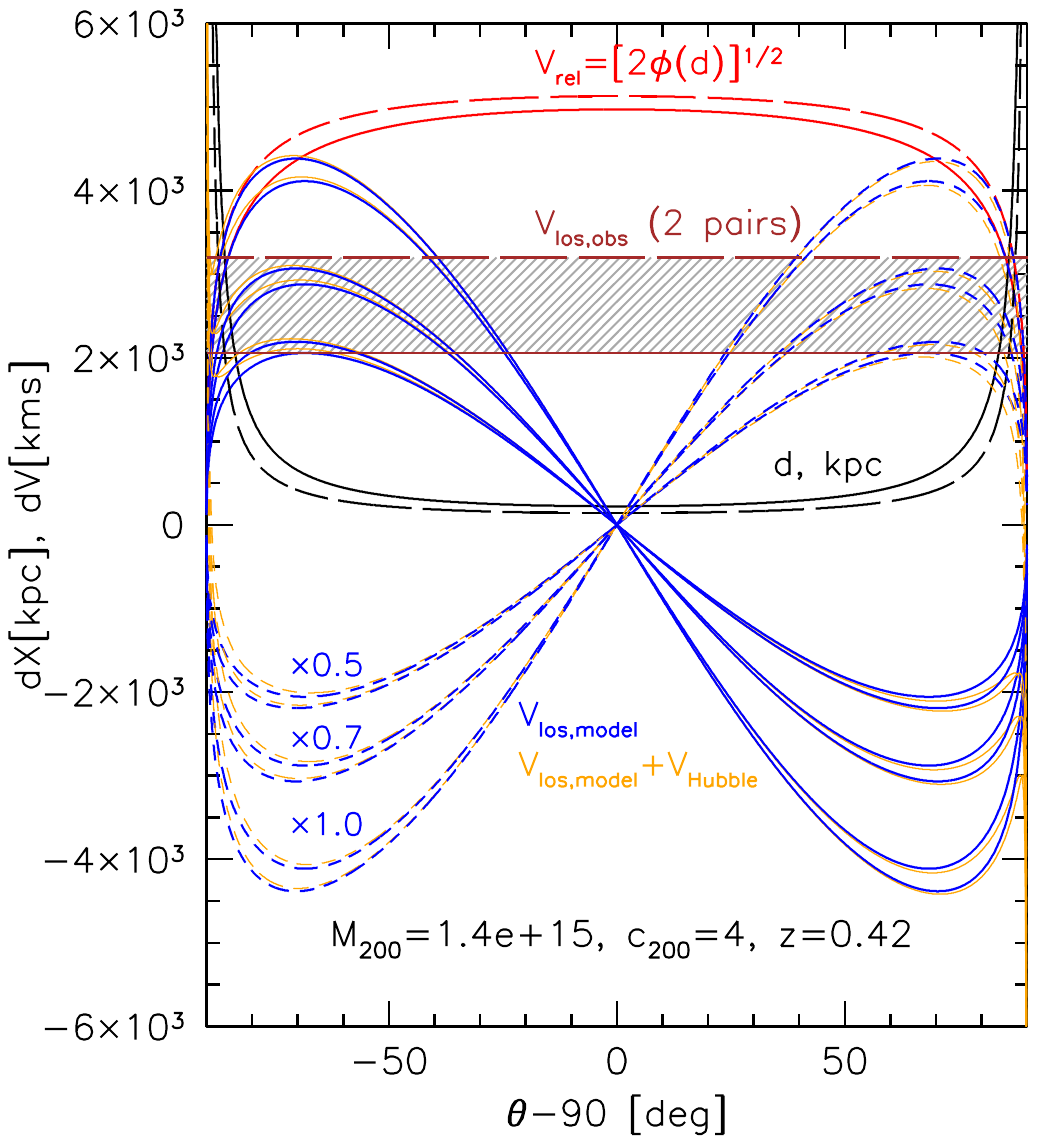}
\caption{
Constraints on the merger angle $\theta$ relative to the line of sight assuming a head-on collision of two subhalos. {\bf Left:} Four possible configurations with the same angle $\theta$ and projected distance $d_{\rm proj}$ between subhalos. Only one of these configurations (marked with a box) fits the scenario discussed here. Namely, (i) the Northern subhalo is receding from us with higher velocity than the Southern one and (ii) this is a "post-merger" phase. {\bf Right:} Expected and observed velocity differences for two pairs of galaxies, which are considered as proxies for subhalo positions and velocities. 
These two pairs have $\Delta V$ and $d_{\rm proj}$ of 2060~${\rm km\,s^{-1}}$ \& 220~kpc and 3200~${\rm km\,s^{-1}}$ \& 142~kpc, respectively. The black and red curves (solid and dashed) show, for each pair, the 3D distance between the galaxies ($d=d_{\rm proj}/\sin\theta$) and the corresponding value of $V_{\rm rel}(d)$ (as given by eq.~\ref{eq:Vesc}). The three close pairs of blue curves show the expected line-of-sight velocity differences $f\times \left( 2\Phi(d)\right)^{1/2}\cos\theta$, where $f$ is the suppression factor discussed in the text. The curves are shown for $f=1, 0.7, 0.5$, where 0.7 is the fiducial value. The almost symmetric "butterfly" pattern of blue curves reflects different geometries shown in the left panel. Only the top-left part of the plot corresponds to the merger scenario discussed here. The thin gold lines show the changes in these curves due to the Hubble expansion, which might be important only in the unlikely case of a merger axis almost perfectly aligned with the line of sight. The two brown horizontal lines (solid and dashed) show the observed line-of-sight velocity difference for each pair. In the idealized case, when both pairs of galaxies perfectly obey eq.~\ref{eq:Vesc} (for a given $f$) and make the same angle to the line of sight, they should intersect two horizontal lines for the same value of $\theta$. However, given the uncertainties of trying to characterize the merger with pairs of galaxies, it is probably sufficient to verify what range of angles is disfavoured by available data. With this in mind, we argue that the merger in the picture plane is definitely excluded (blue curves are very far from the hatched region). A nearly l.o.s. merger might be consistent with data (taking into account Hubble expansion) but the probability of finding such a system is low and the physical distance between galaxies is too large and effectively such halos will be non-interacting objects. 
   }
\label{fig:velocities2}
\end{figure*}

Measured velocities of five galaxies (see Fig.~\ref{fig:z}) corroborate the merger scenario. We tentatively assume that the two galaxies in the North (with $z\sim 0.42$) belong to one subcluster, while the two in the South (with $z\sim 0.41$) belong to the second subcluster. Yet another galaxy to the South from the core has a redshift closer to the redshifts of two galaxies in the Northern clump.  Four possible configurations of the two merging groups are sketched in the left panel of Fig.~\ref{fig:velocities2}. In our favorite scenario, the Northern clump has already passed the pericenter, i.e. the cluster is in the "Post-merger" state. Combined with the larger recession velocity of this clump, the top-left configuration appears the most plausible configuration. 

Given the sparseness of the redshift data, we resort to the simplest estimates of the merger geometry that can relate two observables - the projected distance separation $d_{\rm proj}$ and the line-of-sight velocity difference $V_{\rm los}$. Namely, we consider a test particle radially infalling from infinity into a halo with a given mass $M$ and a Navarro-Frenk-White profile \citep[see e.g.,][for more elaborate analytical or numerical models]{2013ApJ...772..131D, 2019ApJ...881..121W}. 

In this model, the relative velocity $V_{\rm rel}$ of the test particle at separation $d$ from the core of the main cluster is 
\begin{equation}
    V^2_{\rm rel} (d)= 2\: |\Phi^{\rm NFW}(d)| = 2\frac{G M_{200}}{d} \frac{\mathrm{ln}(1+d\, c_{200}/R_{200})}{\mathrm{ln}(1+c_{200})-c_{200}/(1+c_{200})}.
\label{eq:Vesc}
\end{equation}
For $d\rightarrow 0$, $V_{\rm rel}$ is the halo escape velocity $V_{\rm esc}$, which for $c_{200}=4$ is equal to $3.14\sqrt{GM_{200}/R_{200}}$, i.e. $\sim 5500\,{\rm km\,s^{-1}}$ for $M_{200}=1.4\times 10^{15}\,M_\odot$.  
In numerical simulations, the infall velocities of massive halos do not reach values $\approx V_{\rm esc}$, but are systematically lower even when a head-on merger is considered \citep[see, e.g.,][for the discussion of the Bullet cluster case]{2006MNRAS.370L..38H,2007PhRvL..98q1302F,2007MNRAS.380..911S}. The values predicted by eq.~\ref{eq:Vesc} (with $d$ not very close to zero) likely exceed real velocities, too. For instance,  \cite{2023A&A...673A.131L} has recently examined  10 merging clusters in IllustrisTNG300 simulation and derived distances of closest approach and relative velocities. For these mergers, the velocities found in simulations $V_{sim}$ are at the level of  $\sim 0.6 V_{\rm rel}$ (using eq.~\ref{eq:Vesc} for the mass of the main halo and the separation from their Table 1, and setting $c_{200}$ to 4 for all halos). However, for the two most massive mergers, the agreement is better ($V_{\rm sim}\sim\;$0.7 and 0.8$\,V_{\rm rel}$ for these two cases). 
Truncation of the NFW mass distribution at $R_{200}$ \citep[as in the model of][]{2013ApJ...772..131D} reduces the value of $V_{\rm esc}$ by a factor $\sqrt{c_{200}/(1+c_{200})}\approx 0.89$. A further simplifying assumption that two halos maintain their shapes, reduces $V_{\rm esc}$ down to $0.6-0.7$ of $3.14\sqrt{GM_{200}/R_{200}}$, provided the main halo and the subhalo have the same masses/sizes, since in this case, the velocity is set not by the depth of the main halo potential at $d=0$, but rather it is the mass-weighed mean value. 
Given the above considerations, we will assume that velocities, predicted by eq.~\ref{eq:Vesc}, are likely overestimated  (similarly to $V_{\rm esc}$ for $d=0$) and have to be scaled down by a factor in the range $\sim$0.5-0.7-1 when comparing with observations.

We adopted the mass of our  system $M_{200} \simeq 1.4 \cdot 10^{15}$ $M_{\odot}$ as derived from the  $M_{500}$ estimate obtained in Section~\ref{sec:global},  and the mass-concentration relation from \cite{2019ApJ...871..168D}.
Now, for each angle $\theta$ between the merger axis and our line of sight, from the observed distance $d_{\rm proj}$ between merging subhalos and the measured radial velocity difference,
we calculate a separation $d$ between merging clusters in 3D and the total relative velocity using eq.~\ref{eq:Vesc}
(see Figure~\ref{fig:velocities2}).

One subhalo is most plausibly associated with four (visually) clustered galaxies to the North (see Figure~\ref{fig:z}). The position of the second subhalo is uncertain. We tentatively associate it with a galaxy located within the bright X-ray area to the South, about 
40.3 arcsec away in projection from the galaxy to the North. The line-of-sight velocity difference between these two galaxies is $V_{\rm los,obs} = c\Delta z/(1+z_{\rm cl})  \simeq 2000  $ km s$^{-1}$ and the projected distance is $d_{\rm proj}$ = $40.3$ arcsec = 223 kpc. We also consider another galaxy with $z=0.4104$ as a possible candidate for the second subhalo. For this case, the observed velocity difference is $V_{\rm los,obs} = $ 3120 km s$^{-1}$ and   $d_{\rm proj}$ = 25.6 arcsec = 142 kpc  in projection.

Figure~\ref{fig:velocities2} illustrates the constraints on the possible viewing angle. Four possible merger configurations that have the same angle $\theta$ and projected distance $d_{\rm proj}$ between subhalos are sketched in the left panel. The right panel shows the expected l.o.s. velocity difference for a pair of subhalos with a given $\theta$ and projected distance $d_{\rm proj}$. Namely, we calculate $d=d_{\rm proj}/\sin\theta$ and use the scaled version of eq.~\ref{eq:Vesc} to estimate the 3D velocity of merging subclusters, i.e. $\Delta V=f\times \left [ 2 |\Phi^{\rm NFW}(d)| \right]^{1/2}$ assuming that $M_{200}=1.4\times 10^{15}\,M_\odot$ and $c_{200}=4$. The expected line-of-sight velocity difference is then $V_{\rm los,model}=\Delta V \cos\theta$ that can be compared with observations. To this end, we assume that two pairs of galaxies shown in Fig.~\ref{fig:z} can be used as two independent proxies for merging subhalos and plot corresponding  $V_{\rm los,model}$ curves for these two pairs for the entire range of $\theta$ and three values of $f$. Since projected separations are comparable (and small) for both pairs, the expected  $V_{\rm los,model}$ curves are also very similar (see blue lines in  Fig.~\ref{fig:velocities2}). Guided by the Chandra and optical images (Fig.~\ref{fig:centre}, we assume that the system is in a "post-merger" phase, i.e. the Northern group (marked as "N") is moving away from the other group (more to the South from the cluster centroid). Coupled with the larger recession velocity of group N, only the configuration shown in the upper-left corners of both panels is qualitatively consistent with our scenario.  We can now compare the predicted values of $V_{\rm los,model}$ with the observed velocity differences $V_{\rm los,obs}$ for each pair. The latter values are shown in Fig.~\ref{fig:velocities2} with two brown horizontal lines. Ideally, the blue curves might intersect these lines at the same value of $\theta$. Not surprisingly, this does not happen in Fig.~\ref{fig:velocities2}. Indeed, it would be a remarkable coincidence, given that this comparison is based on the (almost) randomly selected galaxies and other uncertainties in the definition of the model. A more reasonable approach is to identify a range of angles that is unlikely suitable for CL0238. Indeed, for $\theta$ close to 90 degrees, i.e. a merger in the sky plane, the expected l.o.s. velocities are obviously small. E.g. for $f=0.7$, values of $\theta \gtrsim 60-70$~deg are implausible. Smaller angles can not be excluded, except for very small angles, i.e. when the merger direction is almost along the line of sight. In this case, the physical separation will be too large and the N and S groups will not be physically interacting, in contradiction to our assumption. The remaining additional argument is simply the probability of observing a randomly oriented system at a given angle $\theta$. This probability is $\propto \sin \theta$ and, therefore, favors large angles. All these indirect arguments suggest that the merger is at some "large" angle that is not yet excluded by the large values of $V_{\rm los,obs}$. For estimates, we set $\theta\sim 45$~deg so that the velocities in the sky plane are of the same order as the observed l.o.s. velocities. Future observations can help to refine the constraints on this angle.  


With these assumptions,  the projected distance of $\sim$200 kpc translates into a 3D separation of $\sim 300$~kpc, i.e. the two subhalos are very close to each other. Using Eq.~\ref{eq:Vesc} and adopting $f=0.7$ we estimate the relative velocity of $\sim 3500\,{\rm km\,s^{-1}}$, the estimated time for crossing 300~kpc is $\lesssim 0.1\,{\rm Gyr}$. This emphasizes the short lifetime of the observed merger phase in CL0238.

As discussed above, both X-ray and optical mass proxies can be to be biased high for merging clusters.
A global observable such as the integrated Comptonization parameter $Y$ (or the equivalent X-ray mass proxy $Y_X$) has been shown to serve as a more robust mass proxy through the entire duration of a merger event \citep[e.g.,][among others]{2006ApJ...650..128K, 2007MNRAS.380..437P, 2008ApJ...680...17W}. To this end, we note that, from the joint analysis of Planck and ROSAT data, \cite{2019yCat..36260007T} estimated the CL0238 mass as  $M_{500}^{\rm XSZ}=6.5\pm 1.2\,\times 10^{14} M_{\odot}$. Converting this value to $M_{200}$ as $\approx 1.4\times M_{500}\approx 9\,10^{14}\, M_{\odot}$ reduces the characteristic velocities by $\sim 15$\%.  This would further shrink the range of plausible angles (and introduce more tension for the pair of galaxies with the l.o.s. velocity difference of 3120~${\rm km\,s^{-1}}$).

%

\section{Discussion}

\subsection{Comparison of CL0238 to MACS0416 and Bullet cluster}

We now proceed with a qualitative comparison of CL0238 with two other prominent merger systems:  MACS J0416.1-2403 and the Bullet cluster.

MACS J0416.1-2403 is a merging cluster at $z=0.396$. Similarly to CL0238, it is characterized by high mass and X-ray temperature ($kT \simeq 10 $ keV, $M\simeq 10^{15} M_{\odot}$; \cite{2015MNRAS.446.4132J,Ogrean2015}), an elongated
mass distribution, and a double-peaked X-ray image \citep{Mann2012}.
Most recent studies favor a pre-collision scenario \citep{Ogrean2015, 2016ApJS..224...33B, 2017ApJ...842..132B} for MACS J0416.1-2403 and confirm that there is no significant offset between the dark matter and the stellar/gas components \citep[see also][for alternative scenarios]{2015MNRAS.446.4132J, 2015MNRAS.447.3130D}. 



The Bullet cluster  (1E 0657-56) at $z=0.296$ is a canonical example of a post-merger cluster. It exhibits spatial offsets between its dark matter and baryonic density peaks derived from the gravitational lensing maps and the gas distribution \citep[see, e.g.,][]{2004ApJ...606..819M,2006A&A...451..395C,2016A&A...594A.121P}. It is also a very massive $M\sim 10^{15} M_{\odot}$ and hot cluster with a spectacular bullet-like X-ray core and a leading bow shock.    

\begin{figure*}
\begin{center}
\includegraphics[trim=0.5cm 0.5cm 0.5cm 0.5cm,clip=true,angle=0,width=1.99\columnwidth]{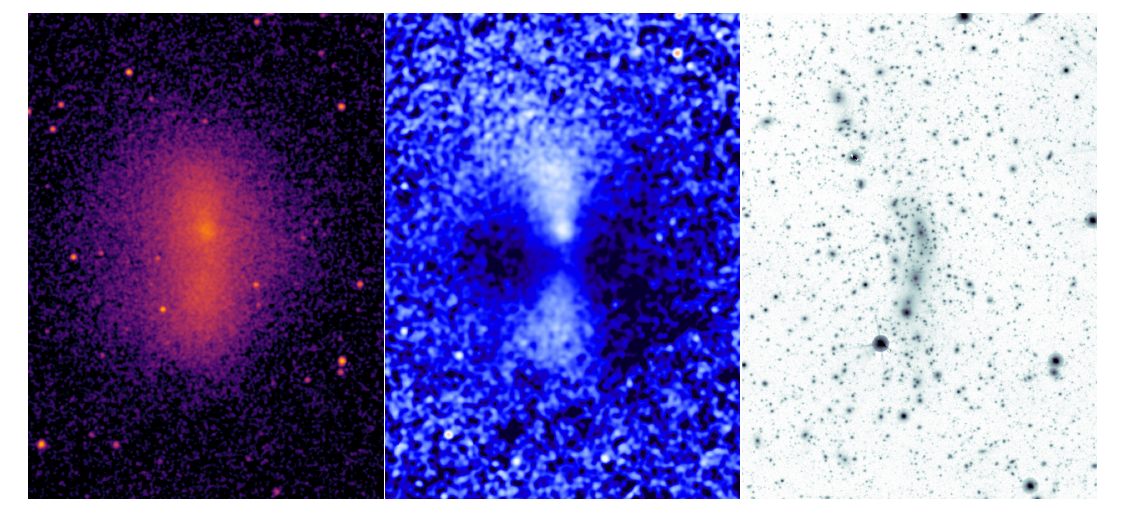}
\includegraphics[trim=0.5cm 0.5cm 0.5cm 0.5cm,clip=true,angle=0,width=1.99\columnwidth]{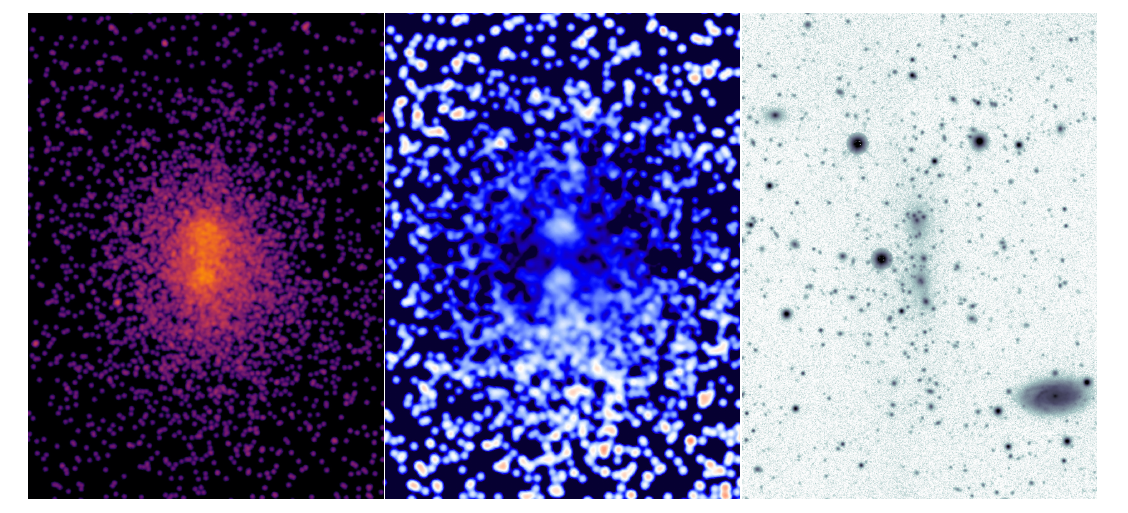}
\includegraphics[trim=0.5cm 0.5cm 0.5cm 0.5cm,clip=true,angle=0,width=1.99\columnwidth]{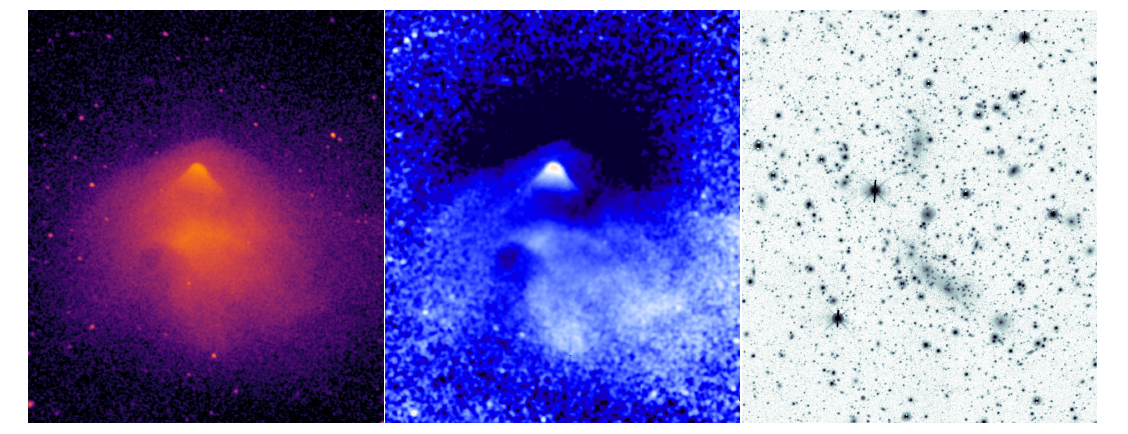}
\end{center}
\caption{Three stages of a merger illustrated by X-ray and optical images of three massive clusters. From top to bottom: before a pericenter passage (MACS J0416), shortly after pericenter passage (CL0238), and a bit later (Bullet). The left panels show Chandra X-ray images.  The panels in the middle present residuals after subtracting, from the X-ray images, the best-fitting symmetric model of a cluster+background and dividing by the model (i.e.  $\displaystyle \frac{\mathrm{image - model}}{\mathrm{model}}$. The right panels show optical DESI r-band images.}
\label{fig:3pigs}
\end{figure*}

How different is CL0238 from MACS J0416.1-2403 and the Bullet? In Figure~\ref{fig:3pigs}, X-ray and optical images of these three clusters are shown (all rotated to align the apparent elongation along the vertical axis). The left and right panels are the Chandra X-ray images and the DESI r-band images (showing the distribution of galaxies), respectively. To obtain images in the middle panels, we first approximate individual X-ray images with a $\beta$-model plus a constant background. Then, from each Chandra image, we subtract the corresponding model and divide by the model (i.e. (image-model)/model). Very prominent deviations from the beta models are seen for all three clusters. 
For MACS0416, we see triangular-shaped features that ``point'' toward each other, hinting at a pre-core-passage phase. 

The Bullet cluster also has two prominent substructures but unlike the MACS0416 cluster, the ``triangle'' now points away from the second substructure with a rather complicated morphology. Such a configuration is robustly identified as a post-merger cluster. 
It is believed that in the Bullet cluster the merger is nearly in the sky plane. The estimates of the relative velocities of its two subclusters come predominantly from the gas density, temperature, and pressure jumps at the prominent bow shock ahead of one of the subclusters. The corresponding estimates of the bow shock Mach number vary between $\sim$2.5 and $\sim3$
\citep{2002ApJ...567L..27M,2004ApJ...606..819M,2006ESASP.604..723M,2008ApJ...679.1173R,2019A&A...628A.100D} implying the shock velocity $\gtrsim  4000\,{\rm km\, s^{-1}}$. However, the velocity of the bow shock can be larger or smaller than that of the galaxies \citep[e.g.][]{2007MNRAS.380..911S,2019MNRAS.482...20Z}. 


CL0238 (middle row) features two peaks in the X-ray image, too. But unlike MACS0416, the optical galaxies (Northern clump) are observed at a larger distance from the center than the nearest X-ray peak. Assuming that ram pressure acting on the gas is responsible for the displacement of the X-ray peaks, we conclude that CL0238 is in the post-pericenter-passage phase and the gas has been strongly shifted from the potential wells traced by galaxies. This suggests a closer analogy to the Bullet cluster.  
However, the apparent (and physical) distance between the peaks is significantly smaller than in the Bullet cluster. We, therefore, propose that CL0238 corresponds to some intermediate phase between the MACS0416 and the Bullet clusters.

Moderately strong bow shocks, expected in the scenario of a close-to-pericenter merger ahead of the subhalos that manage to retain some gas \citep[see][]{2007MNRAS.380..911S,2019MNRAS.482...20Z,2018ApJS..234....4Z}, are not convincingly identified in the relatively shallow X-ray image of CL0238, although a ``bow'' to the South from the blobs is an interesting candidate. Another possibility is that the X-ray emission from the Northern blob is dominated by shocked gas. With deeper X-ray observations one could not only detect a bow shock but also estimate the merger velocity, assuming that the infalling subcluster’s velocity is close to the shock velocity. The shock velocity $V_{\rm shock}$, in turn, is related to the downstream temperature via 
\begin{eqnarray}
    kT=\frac{3}{16}\mu m_p V_{\rm shock}^2\approx 10 \left [\frac{V_{\rm shock}}{2.9\times 10^{3}\,{\rm km\,s^{-1}}}\right ]^2 \,{\rm keV},
\end{eqnarray}
for the gas with the adiabatic index of $5/3$ and small initial temperature, i.e. the case of a strong shock with the Mach number $M\gg 1$. A lower shock velocity will be needed if the gas upstream is hot. For example, for the initial temperature of $kT=6\,{\rm keV}$, a shock velocity of $\sim 2100\,{\rm km\,s^{-1}}$ will be sufficient to have $\sim$10~keV gas downstream.  Of course, for high temperatures and low densities characteristic of CL0238, the electron-ion temperature equilibration time due to Coulomb collisions can be substantial, and for a given shock velocity the electron temperature immediately downstream of the shock might be lower similar to the Bullet cluster \citep[e.g.][]{2006A&A...449..425M}. Useful constraints can be obtained if the temperature, density, and pressure jumps can be measured simultaneously, using X-ray and SZ spatially-resolved data \cite[see ][for the Bullet cluster analysis]{2019A&A...628A.100D}.

\subsection{Lensing}

It is well established that such massive systems as galaxy clusters can act as gravitational lenses revealing properties of high-redshift objects that would otherwise be undetectable with current telescopes (see a recent review by \cite{2024SSRv..220...19N}).
Cluster lenses are now routinely used to study the most distant galaxies \citep[for example,][]{2018MNRAS.476..804D, 2023ApJ...957L..34W}, detect single stars at a cosmological distance \citep[][and others works]{2019ApJ...880...58K}, investigate supernova explosions \citep[for instance,][]{2023ApJ...948...93K, Baklanov2021} and do cosmography with time-delay analyses \citep[][among many others]{2024A&A...684L..23G}.

MACS J0416.1-2403 is one of the  Hubble Frontier Fields clusters and is characterized by a high lensing efficiency \citep{Ogrean2015}. The high lensing efficiency of MACS J0416 is thought to be due to several factors \citep[for details, see][]{Ogrean2015}. First, the high cluster ellipticity increases the ratio of the caustic area in the source plane relative to the critical area, thus generating more multiple images of background sources for the same critical area \citep{Ogrean2015, 2007A&A...461...25M}. Second, ongoing mergers have larger high-magnification regions than non-interacting clusters,  and the total number of multiple images increases with lower separation between merging subhalos, especially when two merging clumps are separated by distance $\le 300$ kpc$/h$ \citep{Ogrean2015, 2004MNRAS.349..476T, 2012A&A...547A..66R}. Another factor driving the lensing efficiency could be the amount of substructures (galaxies) in halos, especially in their central regions. The overall mass profile becomes shallower and provides higher magnifications. Indeed, \cite{2007A&A...461...25M} found that, on average, ellipticity and substructures influence the lensing efficiency almost to the same extent. However, the authors note that the effect of substructures is less important in highly asymmetric lenses.    

As mentioned above, the properties of MACS J0416 are similar to those of CL0238 in many respects. They are located at similar redshifts and have comparable hot gas temperatures and total masses, both clusters are highly elongated as traced by X-ray and optical images. So we suggest that CL0238 is an interesting target for deeper optical+spectroscopic observations and gravitational lensing analysis. Potentially, the lensing efficiency of CL0238 is as high as for MACS J0416, and CL0238 may provide a number of magnified high-redshift sources.

\subsection{SZ, kSZ, and polarization}
A combination of X-ray and SZ data can be used to infer the gas temperature (see Section 4 in \cite{2021A&A...651A..41C} or \cite{2019SSRv..215...17M} for a recent review). To this end, we used the parameters derived from X-ray images and spectra. Namely, the approximation of the radial profile with the simple $\beta$ model yields $r_c=0.47'$, $\beta=0.69$, and a central electron density $n_e=2\times 10^{-3}\,{\rm cm^{-3}}$. For $kT=10\,{\rm keV}$, the expected mean value of the Compton $y=2.5\times 10^{-6}$ within a $10'$ (radius) circle centered at CL0238. 
This value is in good agreement with the mean $y$ extracted from the Planck data $y_{\rm Pl}=2.4\times 10^{-6}$ for a similar region. For this exercise, the PR2 $y$ map was used \citep{2016A&A...594A..22P}. 

\begin{figure}

\includegraphics[angle=0,clip,trim=0cm 4cm 0cm 2cm,clip,width=0.95\columnwidth]{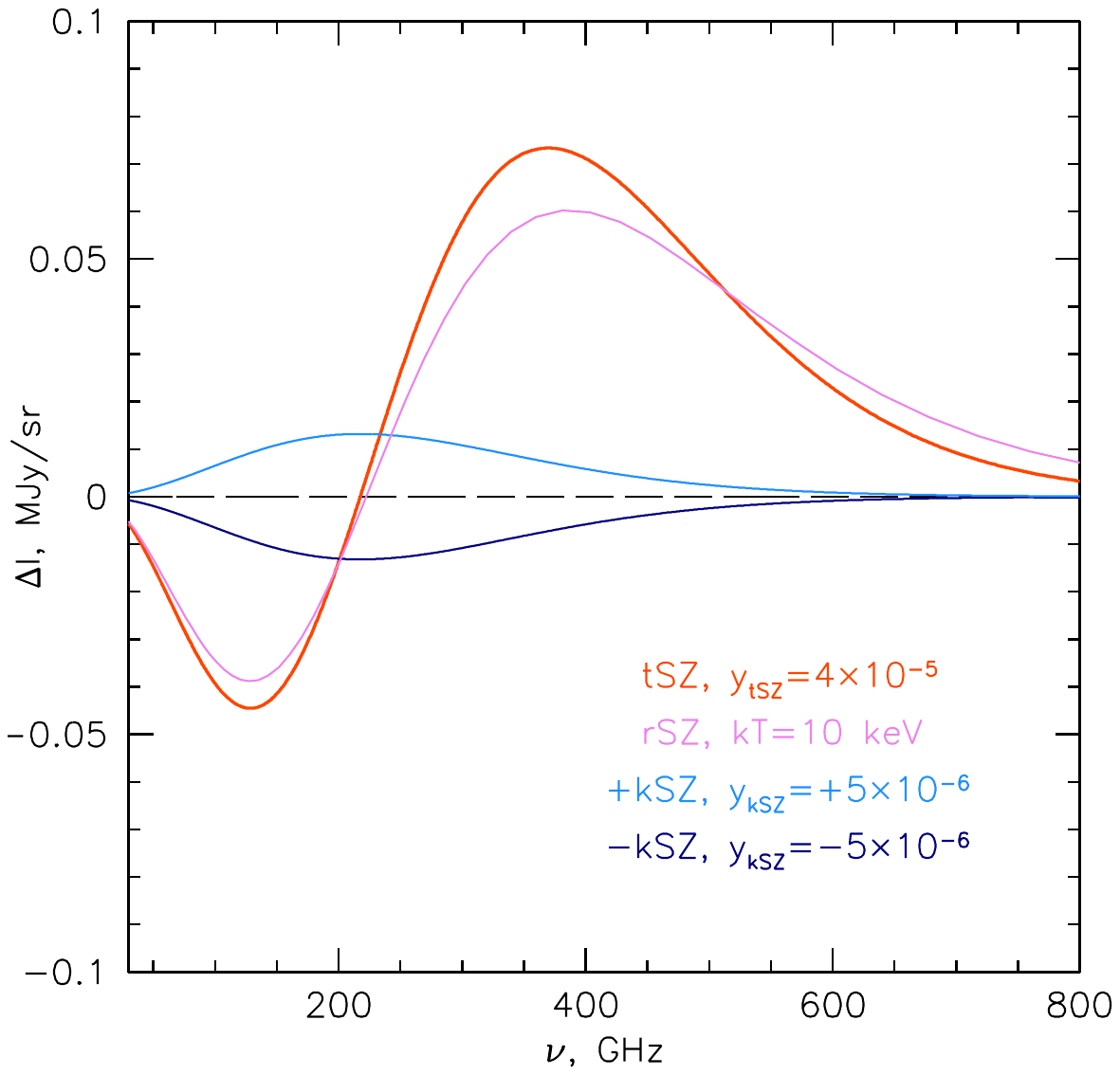}
\caption{Expected SZ signals for the line of sight going through the cluster center. For tSZ (the red curve),  $\tau_T=1.9\times 10^{-3}$ and $kT=10\,{\rm keV}$ are assumed. The kSZ signal (two blue curves) corresponds to the velocity of $\pm 1500\,{\rm km\,s^{-1}}$ and the same $\tau_T$ as for tSZ. For completeness, the magenta line shows the tSZ signal with relativistic corrections for 10~keV electrons.}
\label{fig:tkrsz}
\end{figure}

Given the estimated parameters of the merger, one can expect a substantial contribution of kinematic SZ signal (due to large line-of-sight velocity) \citep{1980MNRAS.190..413S} and polarization in the microwave band (due to large transverse velocity) \citep[][for a review]{1980MNRAS.190..413S, 1999MNRAS.310..765S, 2019SSRv..215...17M}. In particular, Fig.~\ref{fig:tkrsz} shows the estimated tSZ and kSZ signals (in MJy/sr) for the cluster core. The same $\beta$ model was used to estimate $\tau_T$ and $y$ for the cluster core. For kSZ, the amplitudes for $\pm 1500\,{\rm km\,s^{-1}}$ are shown. 

For merger angles $\sim 45^\circ$, similar velocities in the sky plane are expected. In this case, a polarized signal associated with the (sub-)clusters' motion relative to the CMB frame is expected. Its magnitude \citep{1980MNRAS.190..413S} is $\propto \frac{1}{10}\tau_T (\varv_\perp/c)^2\sim 5\times 10^{-9}$ for $\varv_\perp\sim 1500\,{\rm km\,s^{-1}}$.


The caveat of the above estimates is the uncertainty of the gas velocities. Indeed, for a dissociative merger, the gas velocity can be substantially smaller than the velocities of galaxies and/or dark matter. However, if the halos are able to retain the gas after the pericenter passage, the gas can move even faster than the parent halo \citep[a slingshot effect, see, e.g.][]{2007MNRAS.380..911S,2019MNRAS.482...20Z,2019MNRAS.485.2922L,2019ApJ...874..112S}. By combining the optical, X-ray, and kSZ data one can reveal the true geometry of the merger and the merger phase. To this end, additional spectroscopic redshifts, and X-ray data (including XRISM) would be especially useful in combination with spatially resolved SZ data. For the latter, resolving two cores, separated by $\sim 10''$ arcseconds, would require instruments with an angular resolution of $10-20''$ or better, e.g., ALMA, MUSTANG-2, NIKA. 

\subsection{Radio properties of CL0238}

Given the high velocity of the merger, it is plausible that CL0238 might feature non-thermal radio emission, associated with shocks and or radio (mini)halos perturbed by the ongoing merger. The detection of the former would help improve the geometrical model of the merger, while the latter might serve as "tracers" of the gas displaced/stripped from the cores.

\label{sec:radio}

\begin{figure*}
    \centering
    \includegraphics[width=0.95\linewidth]{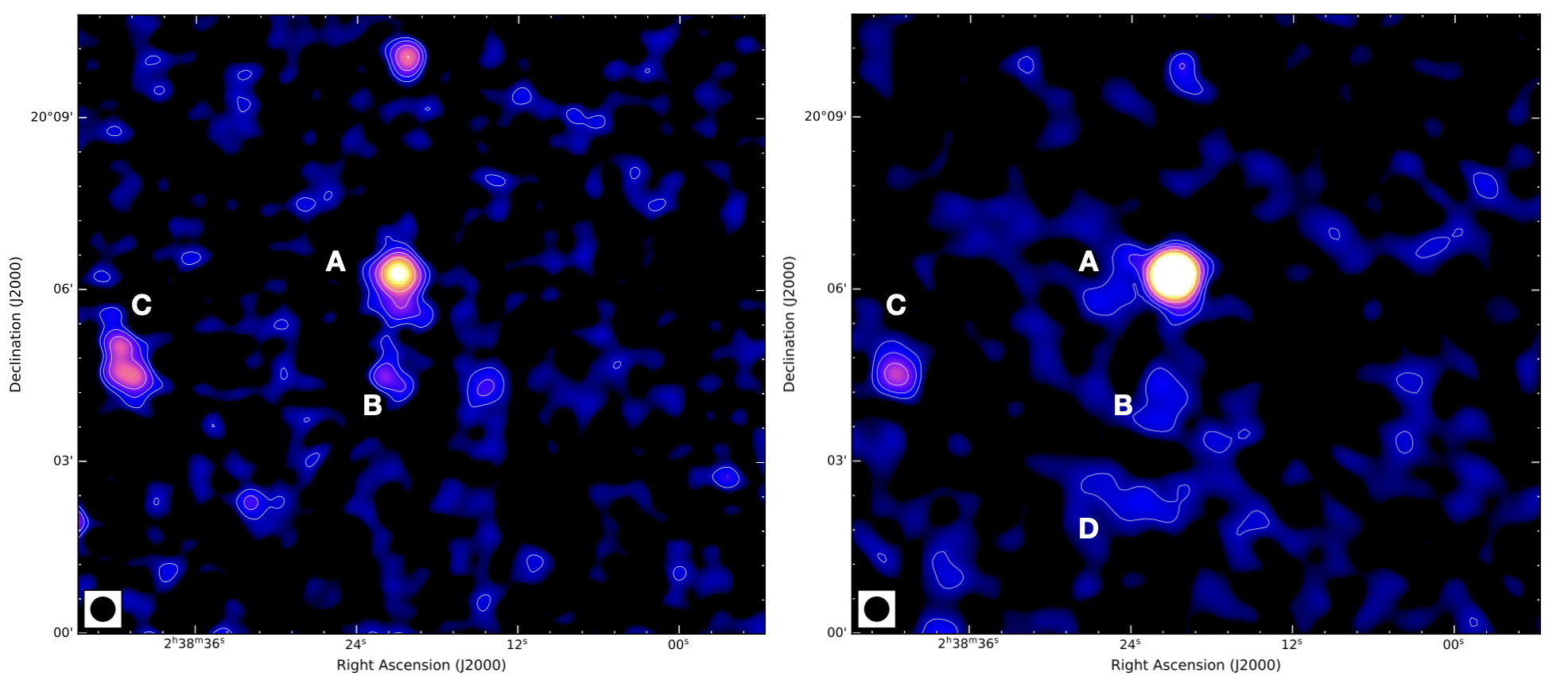}
    \caption{Radio emission from the cluster CL0238 at 887\,MHz (left) and 150\,MHz (right). The 887 MHz and 150\,MHz images are from the RACS and TGSS surveys. In both maps radio contours  level are drawn at  $[1, 2, 4, 8 ...]\times 3.0\sigma_{\rm rms}$. The noise level in the ASKAP and GMRT images are $\rm 170\,\mu Jy\,beam^{-1}$ and $\rm 1.6 \,mJy\, beam^{-1}$, respectively.}
    \label{fig:radio}
\end{figure*}

In Figure \ref{fig:radio}, we show the RACS \citep{McConnell2020} and TGSS \citep[TIFR GMRT Sky Survey;][]{Intema2017} radio images of the cluster at 887\,MHz and 150\,MHz, respectively. The most prominent sources are labeled. Source A is located to the north of the cluster. It has a compact core but there is also a fainter radio emission that apparently extends in the north-south direction. Additionally, there is a hint that source A and source B are connected via a fainter emission (at 2 $\sigma$ level). 

\begin{figure*}
    \centering
    \includegraphics[width=0.45\linewidth]{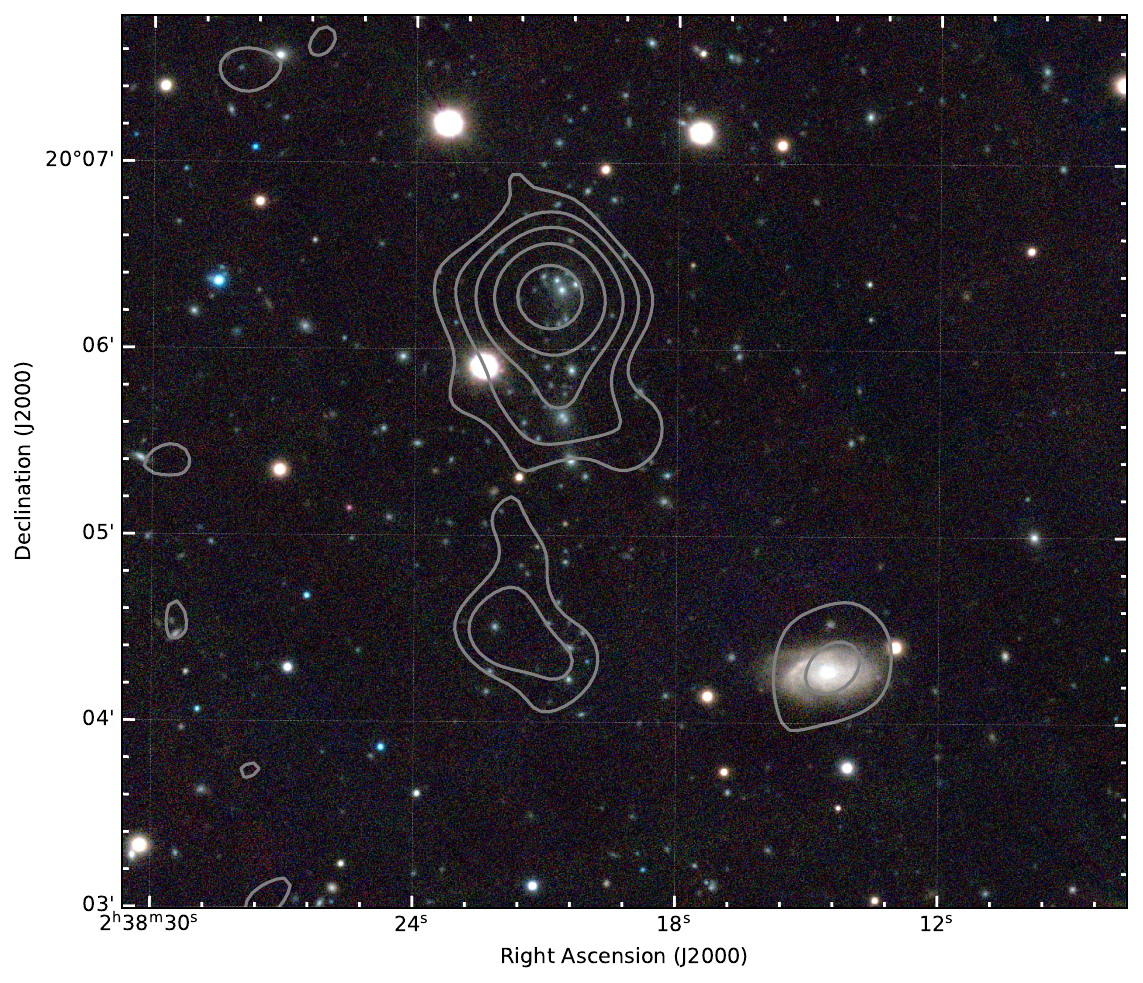}
        \includegraphics[width=0.45\linewidth]{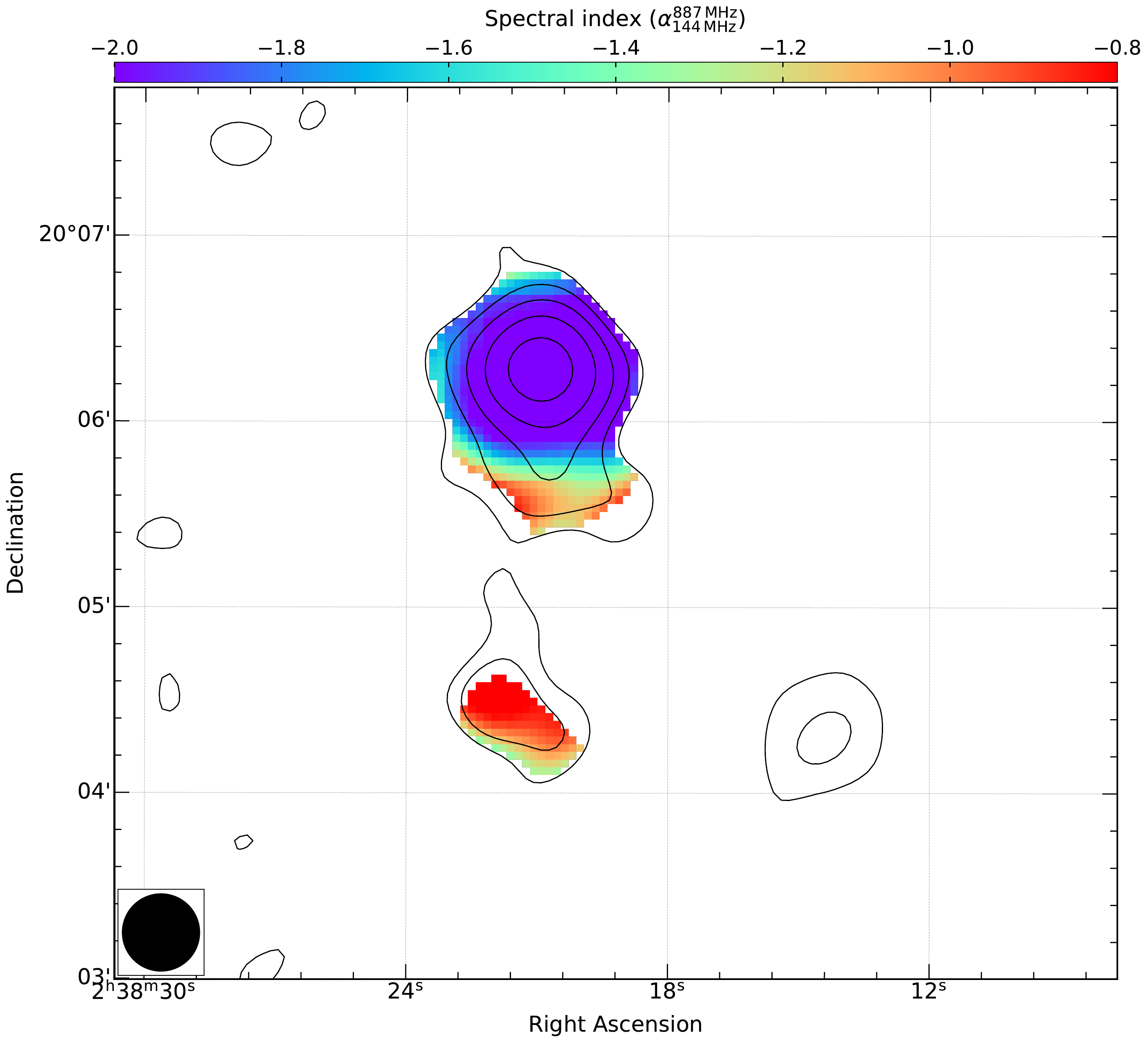}
    \caption{\textit{Left}: Optical image overlaid with the RACS radio contours. \textit{Right:} Spectral index map created between 150 and 887 MHz. Stokes I 887\,MHz radio contours drawn at  $[1, 2, 4, 8 ...]\times 3.0\sigma_{\rm rms}$.}
    \label{fig:radio_index}
\end{figure*}

From the RACS map, we measure that Source A has an extent of about 650\,kpc. It has a flux density of $19\pm2$ mJy, $750\pm150$ mJy, and $1.6\pm0.3$\,Jy at 887, 150 and 88\,MHz (using GLEAM survey), respectively. This indicates that source A has a very steep radio spectral index  \footnote{We define the radio spectral index, $\alpha$, so that $S_{\nu}\propto\nu^{\alpha}$, where $S$ is the flux density at frequency $\nu$} of $\alpha_{88\,\rm MHz}^{887\,\rm MHz}=-1.9\pm0.1$. Optical image at the position of source A is shown in Figure\, \ref{fig:radio_index}. We cannot unambiguously identify an optical counterpart that coincides with the peak radio flux of Source A. The spectral index distribution across it (right panel of Figure \ref{fig:radio_index}) is inconsistent with that of a head-tail radio galaxy, where we expect a flat spectral index in the core region ($-0.5$ to $-0.7$) and gradual spectral steepening across its tails. Moreover, the extremely steep spectral index rules out an identification as a radio loud AGN within the cluster, which typically has a spectral index of $-0.7$. 

Due to its very steep spectral index, location, and small size, Source A could be a radio phoenix. These sources are believed to trace fossil plasma from radio galaxies that have been re-energized by adiabatic compression after the passage of a shock wave in the ICM \citep[e.g.,][]{Ensslin2002, Mandal2019,2019MNRAS.488.5259Z}. However, due to the poor resolution and sensitivity of the existing radio data, its nature remains uncertain.

Source B has an optical counterpart but its redshift is unknown. To the east of the cluster is an arc-like, diffuse, extended source, Source C (see Figure \ref{fig:radio} left panel). It has an extent of about 640\,kpc. The source is partially detected in the 150\,MHz TGSS survey. 

Given the available data, we conclude that there are hints of a displacement and extended features in the data, but high sensitivity data, especially at low frequencies, are needed to draw firm conclusions





\section{Conclusions}

X-ray, SZ, and optical observations show that CL0238 is a massive ($M_{200}\sim 10^{15}M_\odot$) galaxy cluster currently undergoing a merger. At X-ray and optical wavelengths, the cluster is elongated along the North-South direction which is presumably the axis of the merger. Peaks in X-ray and galaxy number density distributions are displaced indicating that the cluster is caught soon after the first core passage. The mean gas temperature is $\sim 10\,{\rm keV}$. There are hints of cooler gas between the two X-ray peaks but more data are needed to confirm its presence. An approximate modeling of the merger geometry (based on a few available spectroscopic redshifts and the spatial distribution of galaxies) suggests that the merger axis makes a substantial angle to the line of sight and, therefore, the true 3D separation of the likely merging subcluster cores is not much larger than their separation in the sky plane, $\sim 200$~kpc. Our approximate modeling also implies that the pericenter passage happened $\lesssim 0.1\,{\rm Gyr}$ ago. 

We compare CL0238 with two well-known clusters MACS0416 and Bullet, and conclude that CL0238 corresponds to an intermediate phase between the pre-merging MACS0416 cluster and the post-merger Bullet cluster. 
The large mass, high elongation, and intermediate redshift of CL0238 make this cluster an interesting target for gravitational lensing analysis. CL0238 is similar in many properties to MACS0416 which is known for its high lensing efficiency, and one could expect that, with deeper optical observations, a considerable number of high redshift sources magnified by the cluster lens CL0238 will be discovered.

\begin{acknowledgements}
We thank Klaus Dolag for helpful disccussions. \\

The authors are grateful to TUBITAK, IKI, KFU and the Tatarstan Academy of Sciences for partial support in the use of RTT-150 (Russian-Turkish 1.5-m telescope in Antalya). The work of IFB, IMH, MAG, MVS, RIG, NAS was supported by the subsidy from the Ministry of Science and Higher Education  of the Russian Federation FZSM-2023-0015, allocated to Kazan Federal University to fulfill the state assignment in the field of scientific activity. \\
IK acknowledges support by the COMPLEX project from the European Research Council (ERC) under the European Union’s Horizon 2020 research and innovation program grant agreement ERC-2019-AdG 882679.\\
WF, CJ, and RK acknowledge support from the Smithsonian Institution, the Chandra
High Resolution Camera Project through NASA contract NAS8-0306, NASA
Grant 80NSSC19K0116 and Chandra Grant GO1-22132X.
\\

In this work, observations with the eROSITA telescope onboard \textit{SRG} space observatory were used. The \textit{SRG} observatory was built by Roskosmos in the interests of the Russian Academy of Sciences represented by its Space Research Institute (IKI) in the framework of the Russian Federal Space Program, with the participation of the Deutsches Zentrum für Luft- und Raumfahrt (DLR). The eROSITA X-ray telescope was built by a consortium of German Institutes led by MPE, and supported by DLR. The SRG spacecraft was designed, built, launched, and operated by the Lavochkin Association and its subcontractors. The science data are downlinked via the Deep Space Network Antennae in Bear Lakes, Ussurijsk, and Baikonur, funded by Roskosmos. The eROSITA data used in this work were converted to calibrated event lists using the eSASS software system developed by the German eROSITA Consortium and analysed using proprietary data reduction software developed by the Russian eROSITA Consortium. \\


This work is based on publicly available optical data from the DESI Legacy Imaging Surveys.
The Legacy Surveys consist of three individual and complementary projects: the Dark Energy Camera Legacy Survey (DECaLS; Proposal ID \#2014B-0404; PIs: David Schlegel and Arjun Dey), the Beijing-Arizona Sky Survey (BASS; NOAO Prop. ID \#2015A-0801; PIs: Zhou Xu and Xiaohui Fan), and the Mayall z-band Legacy Survey (MzLS; Prop. ID \#2016A-0453; PI: Arjun Dey). DECaLS, BASS and MzLS together include data obtained, respectively, at the Blanco telescope, Cerro Tololo Inter-American Observatory, NSF’s NOIRLab; the Bok telescope, Steward Observatory, University of Arizona; and the Mayall telescope, Kitt Peak National Observatory, NOIRLab. Pipeline processing and analyses of the data were supported by NOIRLab and the Lawrence Berkeley National Laboratory (LBNL). The Legacy Surveys project is honored to be permitted to conduct astronomical research on Iolkam Du’ag (Kitt Peak), a mountain with particular significance to the Tohono O’odham Nation.

NOIRLab is operated by the Association of Universities for Research in Astronomy (AURA) under a cooperative agreement with the National Science Foundation. LBNL is managed by the Regents of the University of California under contract to the U.S. Department of Energy.

This project used data obtained with the Dark Energy Camera (DECam), which was constructed by the Dark Energy Survey (DES) collaboration. Funding for the DES Projects has been provided by the U.S. Department of Energy, the U.S. National Science Foundation, the Ministry of Science and Education of Spain, the Science and Technology Facilities Council of the United Kingdom, the Higher Education Funding Council for England, the National Center for Supercomputing Applications at the University of Illinois at Urbana-Champaign, the Kavli Institute of Cosmological Physics at the University of Chicago, Center for Cosmology and Astro-Particle Physics at the Ohio State University, the Mitchell Institute for Fundamental Physics and Astronomy at Texas A\&M University, Financiadora de Estudos e Projetos, Fundacao Carlos Chagas Filho de Amparo, Financiadora de Estudos e Projetos, Fundacao Carlos Chagas Filho de Amparo a Pesquisa do Estado do Rio de Janeiro, Conselho Nacional de Desenvolvimento Cientifico e Tecnologico and the Ministerio da Ciencia, Tecnologia e Inovacao, the Deutsche Forschungsgemeinschaft and the Collaborating Institutions in the Dark Energy Survey. The Collaborating Institutions are Argonne National Laboratory, the University of California at Santa Cruz, the University of Cambridge, Centro de Investigaciones Energeticas, Medioambientales y Tecnologicas-Madrid, the University of Chicago, University College London, the DES-Brazil Consortium, the University of Edinburgh, the Eidgenossische Technische Hochschule (ETH) Zurich, Fermi National Accelerator Laboratory, the University of Illinois at Urbana-Champaign, the Institut de Ciencies de l'Espai (IEEC/CSIC), the Institut de Fisica d’Altes Energies, Lawrence Berkeley National Laboratory, the Ludwig Maximilians Universitat Munchen and the associated Excellence Cluster Universe, the University of Michigan, NSF’s NOIRLab, the University of Nottingham, the Ohio State University, the University of Pennsylvania, the University of Portsmouth, SLAC National Accelerator Laboratory, Stanford University, the University of Sussex, and Texas A\&M University.

BASS is a key project of the Telescope Access Program (TAP), which has been funded by the National Astronomical Observatories of China, the Chinese Academy of Sciences (the Strategic Priority Research Program “The Emergence of Cosmological Structures” Grant \# XDB09000000), and the Special Fund for Astronomy from the Ministry of Finance. The BASS is also supported by the External Cooperation Program of Chinese Academy of Sciences (Grant \# 114A11KYSB20160057), and Chinese National Natural Science Foundation (Grant \# 12120101003, \# 11433005).

The Legacy Survey team makes use of data products from the Near-Earth Object Wide-field Infrared Survey Explorer (NEOWISE), which is a project of the Jet Propulsion Laboratory/California Institute of Technology. NEOWISE is funded by the National Aeronautics and Space Administration.

The Legacy Surveys imaging of the DESI footprint is supported by the Director, Office of Science, Office of High Energy Physics of the U.S. Department of Energy under Contract No. DE-AC02-05CH1123, by the National Energy Research Scientific Computing Center, a DOE Office of Science User Facility under the same contract; and by the U.S. National Science Foundation, Division of Astronomical Sciences under Contract No. AST-0950945 to NOAO.\\

This scientific work uses Rapid ASKAP Continuum Survey (RACS) data obtained from Inyarrimanha Ilgari Bundara/the Murchison Radio-astronomy Observatory. We acknowledge the Wajarri Yamaji People as the Traditional Owners and native title holders of the Observatory site. CSIRO’s ASKAP radio telescope is part of the Australia Telescope National Facility (\url{https://ror.org/05qajvd42}). Operation of ASKAP is funded by the Australian Government with support from the National Collaborative Research Infrastructure Strategy. ASKAP uses the resources of the Pawsey Supercomputing Research Centre. Establishment of ASKAP, Inyarrimanha Ilgari Bundara, the CSIRO Murchison Radio-astronomy Observatory and the Pawsey Supercomputing Research Centre are initiatives of the Australian Government, with support from the Government of Western Australia and the Science and Industry Endowment Fund. This paper includes archived data obtained through the CSIRO ASKAP Science Data Archive, CASDA (\url{https://data.csiro.au}).

\end{acknowledgements}


\bibliographystyle{aa}

\bibliography{example} 

\begin{appendix}


\end{appendix}

\end{document}